\documentclass[12pt,a4paper,titlepage]{article}
\usepackage{epsfig}
\usepackage{amssymb}
\usepackage{amsmath}

\newcommand{\prob}{{\mathbb P}}
\newcommand{\expn}{{\mathbb E}}
\newcommand{\tp}{^{\rm T}}

\newtheorem{thm}{Theorem}
\newtheorem{coro}{Corollary}

\newtheorem{lem}{Lemma}

\begin{document}
\title{Improving Coverage Accuracy of Block Bootstrap Confidence Intervals}
\author{Stephen M.S. Lee\\ {\small\it e-mail:
smslee@hkusua.hku.hk}  \and P.Y. Lai\\
{\small\it e-mail: pylaipy@graduate.hku.hk}}
\date{\small\it Department of Statistics and Actuarial Science, The University of Hong Kong \\
\small\it Pokfulam Road, Hong Kong }
\maketitle

\begin{center}
\bf{ABSTRACT}
\end{center}
The block bootstrap confidence interval based on dependent data can outperform the computationally more convenient normal approximation
only with non-trivial Studentization which, in the case of complicated statistics, calls for highly specialist treatment. 
We propose two different approaches to improving the accuracy of the block bootstrap confidence interval under very general conditions. The first calibrates the coverage level by iterating the block bootstrap. The second calculates Studentizing factors directly from block bootstrap series and requires no non-trivial analytic treatment. 
Both approaches involve two nested levels of block bootstrap resampling and yield high-order accuracy with simple tuning of block lengths at the two resampling levels. A simulation study is reported to provide empirical support for our theory. 
\thispagestyle{empty}

\newpage
\section{Introduction}
The block bootstrap has been developed as a completely model-free procedure for handling inference problems concerning dependent data. A major criticism that impedes widespread acceptance of the procedure in applications is that it lacks second-order accuracy and that empirical selection of block length is critical yet difficult. Although intensive work has been done on the second issue, remedies thus far proposed for the first drawback are rather restrictive in the sense that they require either non-trivial, and sometimes algebraically formidable, Studentization or assumptions of more stringent model structures. Those well-established techniques, such as the iterative bootstrap and the bootstrap-$t$, designed for enhancing bootstrap accuracy for independent data appear to have lost their appeal in the context of dependent data, because the block bootstrap series typically exhibits undesirable artefacts as a consequence of pasting randomly selected data blocks together. An important question is whether the block bootstrap can be made more accurate, by an order asymptotically as well as for finite samples, without
analytically cumbersome Studentization nor having to confine applications to dependent data generated by specific processes.

We investigate formally the applications of two general resampling-based techniques, namely coverage calibration and bootstrap Studentization, to the block bootstrap 
confidence intervals based on dependent data. 
A novel double bootstrap procedure is proposed for either coverage calibration or bootstrap Studentization to improve coverage accuracy of the block bootstrap beyond the first order. The procedure enables both techniques  to retain the simplicity and generality they have already enjoyed when applied to independent data.

Hall (1985) and K\"{u}nsch (1989) introduce the block bootstrap as a fully nonparametric extension of the bootstrap to handle dependent data. 
Its consistency for distributional estimation is verified by K\"{u}nsch (1989) and Liu and Singh (1992). Lahiri (1992) proves for $m$-dependent data that the block bootstrap distribution of an adjusted Studentized sample mean is accurate to second order. Davison and Hall (1993) achieve similar results by kernel-based Studentization. Hall, Horowitz and Jing (1995), 
G\"{o}tze and K\"{u}nsch (1996) and Zvingelis (2003) sharpen the results by giving explicit orders for the estimation error. 

Variants of the block bootstrap include circular block resampling (Politis and Romano, 1992), the stationary bootstrap (Politis and Romano, 1993), 
the matched-block bootstrap (Carlstein, Do, Hall, Hesterberg and K\"{u}nsch, 1998) and the tapered bootstrap (Paparoditis and Politis, 2001). Lahiri (1999) compares the first two with the block bootstrap and confirms superiority of the latter. Davison and Hall (1993), Choi and Hall (2000) and B\"{u}hlmann (2002)
remark on the distortion of dependence structures in block bootstrap series and, for that reason, express doubt over effectiveness of
coverage calibration by bootstrap iterations.

The subsampling method, as studied by Politis and Romano (1994), 
is more generally applicable than the block bootstrap, but has inferior asymptotic properties: see Hall and Jing (1996) and Bertail (1997). Nonparametric methods more accurate than the block bootstrap have been found under more stringent assumptions on the data generating processes.
Examples include the sieve bootstrap (B\"{u}hlmann, 1997; Choi and Hall, 2000) for linear processes, the Markov bootstrap (Rajarshi, 1990) and the local bootstrap (Paparoditis and Politis, 2002) for Markov processes.

We introduce in Section~2 a double bootstrap procedure for either coverage calibration or Studentization of the overlapping block bootstrap. 
Section~3 establishes asymptotic expansions for the coverage probabilities of both the iterated block bootstrap and Studentized block bootstrap 
confidence intervals under sufficiently general regularity conditions, derives the optimal  second-level block length in relation to the first-level
block length and proves asymptotic superiority of our procedures. Section~4 reports a simulation study which compares our methods with the conventional
block bootstrap and two alternative bootstrap-$t$ approaches. Section~5 concludes our findings.
All technical proofs are given in Appendix~\ref{app1}.

\section{Coverage calibration and Studentization}
\subsection{Block bootstrap confidence interval}
Let ${\cal X}=(X_1,\ldots,X_n)$ be a series of $d$-variate observations from the sequence $\{X_i:-\infty<i<\infty\}$, 
which is a realization of a strictly stationary, discrete-time, stochastic process with finite
mean $\mu=\expn[X_1]$. Denote by $\bar{X}=\sum_{i=1}^nX_i/n$ the sample mean.

We briefly review the block bootstrap construction of a level $\alpha$ upper confidence bound for a scalar parameter of interest $\theta=H(\mu)$, for some smooth function 
$H:{\Bbb R}^d\rightarrow{\Bbb R}$. A natural plug-in estimator of $\theta$ is $\hat\theta=H(\bar{X})$. 
This smooth function model setup encompasses a wide variety of estimators, or their high-order asymptotic approximations, providing a sufficiently general platform for investigating 
the block bootstrap confidence procedure.

For a block length $\ell$ ($1\leq\ell\leq n$), let $n'=n-\ell+1$ and define overlapping blocks 
$Y_{j,\ell}=(X_j,X_{j+1},\ldots,X_{j+\ell-1})$, $j=1,\ldots,n'$. A generic first-level block bootstrap series 
${\cal X}^*=(X^*_1,\ldots,X^*_{b\ell})$, where $b=\langle n/\ell\rangle$ and $\langle x\rangle $ denotes the integer part
of $x$, is given by sampling $b$ blocks randomly with replacement from $\{Y_{j,\ell}:1\leq j\leq n'\}$ and pasting them end-to-end in the order sampled, so that 
$(X^*_{(j-1)\ell+1},\ldots,X^*_{j\ell})$ denotes the $j$th block sampled, $j=1,\ldots,b$. 

Let $\prob^*$ and $\expn^*$ denote the probability measure and expectation operator induced
by block bootstrap sampling, conditional on ${\cal X}$, respectively. Define $\bar{X}^*=\sum_{i=1}^{b\ell}X_i^*/(b\ell)$ and the block bootstrap distribution function
${G}^*(x)=\prob^*\left((b\ell)^{1/2}[H(\bar{X}^*)-H(\expn^*\bar{X}^*)]\leq x\right)$, $x\in{\Bbb R}$.
Then ${\cal I}(\alpha)=\hat\theta-n^{-1/2}{G}^{*-1}(1-\alpha)$ defines a level $\alpha$ block bootstrap upper confidence bound for $\theta$. 
Note that sampling of overlapping blocks incurs an edge effect which explains the use of $H(\expn^*\bar{X}^*)$, rather than the more conventional $\hat\theta=H(\bar{X})$,
for centering the bootstrap estimator in the definition of $G^*$. Under regularity conditions to be detailed in Section~3, the choice $\ell\propto n^{1/3}$ yields the smallest
coverage error, of order $O(n^{-1/3})$, for ${\cal I}(\alpha)$. 

\subsection{Second-level block bootstrap}
For independent and identically distributed data, coverage calibration and Studentization provide two well-known techniques for improving coverage accuracy of bootstrap confidence intervals.
We consider applications of the two techniques in the present context of dependent data. Both coverage calibration and the version of Studentization proposed herein call for a double bootstrap
procedure as described below.

Based on ${\cal X}^*$, define blocks $Y^*_{i,j,k}=(X^*_{(i-1)\ell+j},X^*_{(i-1)\ell+j+1}, 
\ldots,X^*_{(i-1)\ell+j+k-1})$, each of length $k$ ($1\leq k\leq \ell$), for $i=1,\ldots,b$ and  $j=1,\ldots,\ell'$, where $\ell'=\ell-k+1$.
Note that for each fixed $i=1,\ldots,b$, $Y^*_{i,1,k},\ldots,Y^*_{i,\ell',k}$ represent overlapping blocks within the block $(X^*_{(i-1)\ell+1},\ldots,X^*_{i\ell})$, which is itself sampled
randomly from the blocks $\{Y_{j,\ell}:1\leq j\leq n'\}$. 
The second-level block bootstrap series, denoted by ${\cal X}^{**}=(X^{**}_1,\ldots,X^{**}_{ck})$, for $c=\langle n/k\rangle$,
is sampled from the $b\ell'$ blocks $\{Y^*_{i,j,k}:1\leq i\leq b, 1\leq j\leq \ell'\}$ in the same way as is ${\cal X}^*$ from $\{Y_{j,\ell}:1\leq j\leq n'\}$.
That $Y^*_{i,j,k}$ is a subseries of $k$ consecutive observations within ${\cal X}$ eliminates the possibility of drawing second-level blocks that run across joints of 
the first-level block bootstrap series, thereby avoiding the discontinuity problem which has aroused forejudged criticisms about the very usefulness of the double block bootstrap.

Denote by $\prob^{**}$ and $\expn^{**}$ respectively the probability measure and expectation operator induced
by second-level block bootstrap sampling, conditional on ${\cal X}^*$. Define $\bar{X}^{**}=\sum_{i=1}^{ck}X_i^{**}/(ck)$ and
\[{G}^{**}(x)=\prob^{**}\left((ck)^{1/2}[H(\bar{X}^{**})-H(\expn^{**}\bar{X}^{**})]\leq x\right),\;\;\;x\in{\Bbb R}.\]
The second-level block bootstrap distribution $G^{**}$ can be used in two different ways, namely coverage calibration and Studentization, to correct ${\cal I}(\alpha)$:

\noindent
1. {\em Coverage calibration\/} ---

The coverage calibration method adjusts the nominal level $\alpha$ to $\hat\alpha$, obtained as solution to the equation 
\[\prob^*\left((b\ell)^{1/2}[H(\bar{X}^*)-H(\expn^*\bar{X}^*)]\leq G^{**-1}(1-\hat\alpha)\right)=1-\alpha.\]
The coverage-calibrated upper confidence bound is then
${\cal I}_C(\alpha)={\cal I}(\hat\alpha)=\hat\theta-n^{-1/2}{G}^{*-1}(1-\hat\alpha)$.

\noindent 2. {\em Studentization\/} ---

Let $\hat\tau$  be the conditional standard deviation of $(b\ell)^{1/2}H(\bar{X}^*)$ given ${\cal X}$, and $\tau^*$ be that of  $(ck)^{1/2}H(\bar{X}^{**})$ given ${\cal X}^*$. Define, for $x\in{\Bbb R}$,
$J^*(x)=\prob^*\left((b\ell)^{1/2}[H(\bar{X}^*)-H(\expn^*\bar{X}^*)]/\tau^*\leq x\right)$. The level $\alpha$ Studentized upper confidence bound is then given by
${\cal I}_S(\alpha)=\hat\theta-n^{-1/2}\hat\tau{J}^{*-1}(1-\alpha)$.

We show in Section~3 that under
regularity conditions, ${\cal I}_C(\alpha)$ and ${\cal I}_S(\alpha)$ are asymptotically equivalent up to order $O_p\left(k^{-2}n^{-1/2}+\ell n^{-3/2}\right)$.
Both methods enjoy a reduced coverage error of order $O(n^{-2/3})$
if we set, for example, $2k=\ell\propto n^{1/3}$.  
Our results rebut the criticisms expressed by, for example, Davison and Hall (1993), Choi and Hall (2000) and B\"{u}hlmann (2002)
over the effectiveness of coverage calibration. 
Indeed, ${\cal I}_C(\alpha)$ is the first ever non-Studentized block bootstrap interval having the same order of coverage accuracy as has previously been shown to be possible
only with Studentization under the present regularity conditions. This has especially important implications for problems in which Studentization is found to be numerically unstable and
therefore results in highly variable interval endpoints.
On the other hand, construction of ${\cal I}_S(\alpha)$
makes unnecessary all those non-trivial, problem-specific, algebraic manipulations which are instrumental to calculation of the Studentizing factors suggested by
Lahiri (1992), Davison and Hall (1993) and G\"{o}tze and K\"{u}nsch (1996).
Indeed, both $\hat\tau$ and $\tau^*$ are readily obtained by direct Monte Carlo simulation from
the bootstrap distributions $G^*$ and $G^{**}$ respectively, thus adhering most closely to the celebrated plug-in principle underlying the very bootstrap methodology.

\section{Theory}
Higher-order asymptotic investigation of coverage accuracy of the block bootstrap confidence bounds is possible if we assume regularity conditions that facilitate Edgeworth expansions of the distribution functions of $n^{1/2}(\hat\theta-\theta)$ and $n^{1/2}(\hat\theta-\theta)/\hat\tau$.  
The set of conditions considered by G\"{o}tze and Hipp (1983) has generally been accepted as the standard assumptions underpinning a high-order asymptotic theory of the block bootstrap. Importantly, previous studies have shown that the block bootstrap can be made accurate to second order only with non-trivial Studentization or substantial strengthening of the G\"{o}tze and Hipp conditions. We shall establish asymptotic results for our coverage calibration and Studentization approaches under the G\"{o}tze and Hipp conditions, as 
modified by Lahiri (2003, Section~6.5) below, with
$\|\cdot\|$ denoting the usual Euclidean norm:
\begin{itemize}
\item[(A1)] $\expn\|X_1\|^{35+\delta}<\infty$ for some $\delta>0$.
\item[(A2)] $\lim_{n\rightarrow\infty}{\rm Cov}\left(n^{-1/2}\sum_{i=1}^nX_i\right)$ exists and is nonsingular.
\item[(A3)] There exists a constant $C\in(0,1)$ such that for $i,j>1/C$, 
\[\inf\left\{s\tp{\rm Cov}\left(\sum_{r=i+1}^{i+j}X_r\right)s:\|s\|=1\right\}>Cj.\]
\item[(A4)] There exist a constant $C>0$ and sub-$\sigma$-fields ${\cal D}_0,{\cal D}_{\pm 1},\ldots$ of the $\sigma$-field underlying the probability space induced  by $X_1$ such that
for $i,j=1,2,\ldots\:$,
\begin{itemize}
\item[(i)] there exist ${\cal D}^{i+j}_{i-j}$-measurable random vectors $\tilde{X}_{i,j}$ satisfying
$\expn\|X_i-\tilde{X}_{i,j}\|\leq C^{-1}e^{-Cj}$ for $j>1/C$, where ${\cal D}_r^s$ denotes the sigma-field generated by $\{{\cal D}_t:r\leq t\leq s\}$;
\item[(ii)] $|\prob(A\cap B)-\prob(A)\prob(B)|\leq C^{-1}e^{-Cj}$ for any $A\in{\cal D}^i_{-\infty}$ and $B\in{\cal D}^\infty_{i+j}$;
\item[(iii)] $\expn\left|\expn\left[\exp(\iota s\tp\sum_{r=i-j}^{i+j}X_r)\mid\{{\cal D}_t:t\neq i\}\right]\right|\leq e^{-C}$ for $i>j>1/C$ and $s\in{\Bbb R}^d$ with $\|s\|\geq C$,
where $\iota^2=-1$;
\item[(iv)] $\expn\left|\prob(A\mid\{{\cal D}_t:t\neq i\})-\prob(A\mid\{{\cal D}_t:0<|t-i|\leq j+r\})\right|\leq C^{-1}e^{-Cj}$ for $r=1,2,\ldots$ and $A\in{\cal D}_{i-r}^{i+r}$.
\end{itemize}
\end{itemize}
Note that (A4) introduces an auxiliary set of sub-$\sigma$-fields ${\cal D}_t$ to bring a wide variety of weakly dependent processes under a common framework. Special examples include linear processes, $m$-dependent shifts, stationary homogeneous Markov chains and stationary Gaussian processes. 

Bhattacharya and Ghosh's (1978) smooth function model supplies a rich class of estimators and has been extensively studied in the bootstrap literature: see, for example, Hall (1992). In the dependent data context, it encompasses estimators such as sample autocovariances, sample autocorrelation coefficients, sample partial autocorrelation coefficients and Yule-Walker estimators for autoregressive processes. Importantly, the model admits highly-structured asymptotic expansions to facilitate establishment of Edgeworth expansions and their block bootstrap versions. 
We adopt the smooth function model as described by G\"{o}tze and K\"{u}nsch (1996) under the assumption
\begin{itemize}
\item[(A5)] $H:{\Bbb R}^d\rightarrow{\Bbb R}$ is four times continuously differentiable with non-vanishing gradient at $\mu$ and fourth-order derivatives at $x\in{\Bbb R}^d$ bounded in magnitude by
$C(1+\|x\|^D)$ for fixed constants $C,D>0$.
\end{itemize}
Next we introduce some notation. Write  $x=(x^{(1)},\ldots,x^{(d)})$ for each $x\in{\Bbb R}^d$. Define, for $r_1,r_2,\ldots=1,\ldots,d$ and $i_1,i_2,\ldots=0,1,2,\ldots\:$, 
$\gamma^{r_1,r_2,\ldots}_{i_1,i_2,\ldots}=\expn\left[(X_0-\mu)^{(r_1)}(X_{i_1}-\mu)^{(r_2)}(X_{i_2}-\mu)^{(r_3)}\cdots\right]$. 
For $r,s,\ldots=1,\ldots,d$, define $H_r=\left.(\partial/\partial x^{(r)})H(x)\right|_{x=\mu}$, $H_{rs}=\left.(\partial^2/\partial x^{(r)}\partial x^{(s)})H(x)\right|_{x=\mu}$,
etc. Under conditions (A1)--(A4), we can expand the variance-covariance matrix of $S_n=n^{-1/2}\sum_{i=1}^n(X_i-\mu)$ such that
${\rm Cov}\left(S_n^{(r)},S_n^{(s)}\right)=\chi_{2,1}^{r,s}+n^{-1}\chi^{r,s}_{2,2}+O(n^{-2})$, $r,s=1,\ldots,d$,
for constants $\chi_{2,1}^{r,s}$ and $\chi^{r,s}_{2,2}$ not depending on $n$. 
In particular, we have $\chi^{r,s}_{2,1}=\sum_{i=-\infty}^\infty\gamma^{r,s}_i$.
Define $\sigma^2={\rm Var}\left(\sum_{r=1}^dH_rS_n^{(r)}\right)$, which, under the above conditions, 
is positive and has order $O(1)$. Let $\phi(\cdot)$ and
$z_\xi$ be the standard normal density function and $\xi$th quantile respectively.

Our main theorem below derives expansions for the coverage probabilities of the various block bootstrap upper confidence bounds.
\begin{thm}
\label{thm:cov}
Let $\{X_i:-\infty<i<\infty\}$ be a strictly stationary, discrete-time, stochastic process with finite mean $\mu=\expn[X_1]$. Let $\alpha\in(0,1)$ be fixed. 
Assume that conditions (A1)--(A5) hold.
Then, 
\begin{itemize}
\item[(i)] for $\ell=O(n^{1/3})$ and $\ell/n^\epsilon\rightarrow\infty$ for some $\epsilon\in(0,1)$,
\begin{eqnarray}
\lefteqn{\prob(\theta\leq{\cal I}(\alpha))}\nonumber\\
&=&\alpha+\ell^{-1}2^{-1}\sigma^{-2}z_\alpha\phi(z_\alpha)\sum_{r,s=1}^dH_rH_s\chi^{r,s}_{2,2}-n^{-1/2}2^{-1}\sigma^{-3}z^2_\alpha\phi(z_\alpha)\nonumber\\
&&\times\,\left\{\sum_{r,s,t=1}^dH_rH_sH_t\!\!\!\sum_{i,j=-\infty}^\infty\!\!\!\gamma^{r,s,t}_{i,j}
+2\!\!\!\sum_{r,s,t,u=1}^dH_rH_sH_{tu}\chi^{r,t}_{2,1}\chi^{s,u}_{2,1}\right\}\nonumber\\
&&\;\;\;+\,O\left(\ell^{-2}+\ell n^{-1}\right); \label{eq:boot}
\end{eqnarray}
\item[(ii)] for $k\leq\ell=O(n^{1/3})$ and $k/n^\epsilon\rightarrow\infty$ for some $\epsilon\in(0,1)$, the confidence limits
${\cal I}_C(\alpha)$ and ${\cal I}_S(\alpha)$ differ by $O_p\left(k^{-2}n^{-1/2}+\ell n^{-3/2}\right)$ and have
coverage probability
\begin{equation}
\label{eq:it.boot}
\alpha+(2\ell^{-1}-k^{-1})2^{-1}\sigma^{-2}z_\alpha\phi(z_\alpha)\sum_{r,s=1}^dH_rH_s\chi^{r,s}_{2,2}+O\left(k^{-2}+\ell n^{-1}\right).
\end{equation}
\end{itemize}
\end{thm}
It is clear from Theorem~\ref{thm:cov} that ${\cal I}(\alpha)$ has coverage error of order $O(\ell^{-1}+\ell n^{-1})$, which can be reduced by either coverage calibration or
Studentization to  $O(\ell^{-2}+\ell n^{-1})$ if we set $k=\ell/2$.
Heuristically, a chief source of coverage error of ${\cal I}(\alpha)$ stems from the large bias, of order $1/\ell$, of the block bootstrap variance estimator. 
The second-level block bootstrap variance estimator has leading bias of order $1/k-1/\ell$ when viewed as an estimator of the first-level block bootstrap variance estimate. 
Existence of such second-level bias term enables either the coverage calibration or Studentization strategies to automatically offset the first-level bias of order $1/\ell$, 
provided that $k$ is set to $\ell/2$. Furthermore,
expansions (\ref{eq:boot}) and (\ref{eq:it.boot}) enable us to derive the optimal choices of block lengths $\ell$ and $k$ for achieving the best coverage error rates. 
We see from (\ref{eq:boot}) that, in the absence of coverage calibration or Studentization, the optimal block length $\ell$ should have order $n^{1/3}$ in order to yield the smallest coverage error, of order
$O(n^{-1/3})$, for ${\cal I}(\alpha)$. 
With $k=\ell/2$ and $\ell\propto n^{1/3}$, the coverage error of both ${\cal I}_C(\alpha)$ and ${\cal I}_S(\alpha)$
has order $O(n^{-2/3})$, a significant improvement over that of the unmodified ${\cal I}(\alpha)$. The following corollary summarizes the above results.
\begin{coro}
\label{coro:cov}
Under the conditions of Theorem~\ref{thm:cov}, 
\begin{itemize}
\item[(i)] ${\cal I}(\alpha)$ has coverage error of order $O(n^{-1/3})$, achieved by setting $\ell\propto n^{1/3}$;
\item[(ii)] ${\cal I}_C(\alpha)$ and ${\cal I}_S(\alpha)$ are asymptotically equivalent up to order $O_p(n^{-7/6})$ and have coverage error of order $O(n^{-2/3})$, achieved by setting $2k=\ell\propto n^{1/3}$.
\end{itemize}
\end{coro}
Corollary~\ref{coro:cov} confirms that second-order correction of the block bootstrap interval can be achieved by straightforward application of either coverage calibration or Studentization.
Previous approaches proposed in the literature to such second-order correction rely invariably on explicit computation of a non-trivial expression of the Studentizing factor, which must be analytically derived for each smooth function model under study. See, for example, H\"{a}rdle, Horowitz and Kreiss (2003) for a review of such approaches.
At the expense of computational efficiency incurred by the double bootstrap procedure, calculation of ${\cal I}_C(\alpha)$ or ${\cal I}_S(\alpha)$ involves no analytic formula and can be
carried out by brute force Monte Carlo simulation. 
Perhaps surprising is the extremely simple relationship ($k=\ell/2$) between the optimal first-level and 
second-level block lengths, which relieves us of the notoriously difficult task of determining the best block length for the double block bootstrap, in so far as the selection of $k$ is concerned.

\section{\label{sim}{Simulation study}}
We conducted a simulation study to investigate the empirical performance of ${\cal I}_C(\alpha)$ and ${\cal I}_S(\alpha)$ in comparison with ${\cal I}(\alpha)$.
Two other Studentized block bootstrap confidence bounds, based on constructions of Davison and Hall (1993) and G\"{o}tze and K\"{u}nsch (1996) and denoted
by ${\cal I}_{DH}(\alpha)$ and ${\cal I}_{GK}(\alpha)$ respectively,  were also included in the study for 
reference: see Appendix~\ref{app2} for details of these two latter approaches. 
Time series data were generated under the following three models: 
\begin{itemize}
\item[(a)] ARCH(1) process: $X_i=e_i(1+0.3X_{i-1}^2)^{1/2}$, 
\item[(b)] MA(1) process: $X_i=e_i+0.3e_{i-1}$,
\item[(c)] AR(1) process: $X_i=0.3X_{i-1}+e_i$, 
\end{itemize}
where the $e_i$ are independent $N(0,1)$ variables. The parameter $\theta$ was taken to be the mean, variance and lag 1 autocorrelation, and the nominal level $\alpha$ was set to be
$0.05$, 0.10, 0.90 and 0.95. For each method, the coverage probability of the level $\alpha$ upper confidence bound was approximated by averaging over 1000 independent time series 
of length $n=500$ and 1000. Construction of each confidence bound was based on 1000 first-level block bootstrap series using block length $\ell=\langle n^{1/3}\rangle$, in addition to which 
1000 second-level series based on block length $k=\langle \ell/2\rangle$ were generated from each first-level series to construct ${\cal I}_C(\alpha)$ and ${\cal I}_S(\alpha)$.
Specifically, we have $(\ell,k)=(8,4)$ and $(10,5)$ for $n=500$ and 1000 respectively.
The constant $c$ was set to be 0.5 in the calculation of the Studentizing factor for ${\cal I}_{GK}(\alpha)$: see Appendix~\ref{app2}.

The coverage results are given in Tables~\ref{tab:mean}--\ref{tab:corr} for the mean, the variance and the lag 1 autocorrelation cases respectively.
In general, coverage calibration and all three Studentization methods succeed in reducing coverage error of ${\cal I}(\alpha)$ when the latter is noticeably inaccurate such as 
for $\theta={\rm Var}(X_1)$. Our proposed ${\cal I}_C(\alpha)$ and ${\cal I}_S(\alpha)$ either outperform or are comparable to ${\cal I}_{DH}(\alpha)$ and ${\cal I}_{GK}(\alpha)$
in the variance and lag 1 autocorrelation cases. Note that ${\cal I}_{GK}(\alpha)$ is exceptionally poor for small $\alpha$ in the autocorrelation case. 
All five confidence bounds have similar performance when $\theta=\expn[X_1]$.

\section{Conclusion}
We have proposed two double bootstrap approaches, one for calibrating the nominal coverage and the other for calculating the Studentizing factor, to improving accuracy of the block bootstrap 
confidence interval. The main advantage of the proposed approaches lies in the ease with which the second-level block length $k$ can be determined, namely half the first-level block length, and
the Studentizing factor can be computed, essentially by a trivial application of the plug-in principle. Not in the literature has the same degree of improvement been achieved without analytic
derivation of the Studentizing factor in a highly problem-specific manner. The problem of empirical determination of the first-level block length $\ell$ has been dealt with by various 
authors but methods which have proven satisfactory performance are not yet available. 
Both theoretical and empirical findings suggest that our proposed coverage calibration or Studentization approaches are effective in reducing coverage error even in the absence of a
sophisticated data-based scheme for selecting $\ell$ in the confidence procedure. 
While implementation of the approaches is analytically effortless, the only price to pay is the extra computational cost induced by the second level of block bootstrapping.  

Although our focus is confined to the smooth function model setting, it is believed that similar results extend also to von Mises-type functionals as well as to estimating functions, after
appropriate modifications of the proof of our main theorem.
Extension to dependence structures outside the present framework, such as series exhibiting long-range dependence, is less trivial and worth investigating in future studies.

\section{Appendix}
\subsection{\label{app1}Proof of Theorem~\ref{thm:cov}}
We first state a few lemmas concerning moments of centred sums of stationary observations and their bootstrap counterparts.

Define $S^*_n\equiv (b\ell)^{1/2}\left(\bar{X}^*-\expn^*\bar{X}^*\right)$ and $S^{**}_n\equiv (ck)^{1/2}\left(\bar{X}^{**}-\expn^{**}\bar{X}^{**}\right)$.
Define, for $i=0,\pm 1,\ldots$ and $r=1,2,\ldots\:$, $Z_i=X_i-\mu$ and $V_{i,r}=r^{-1/2}\sum_{s=i}^{i+r-1}Z_s$.
\begin{lem}
\label{lem:lahiri}
Under the conditions of Theorem~\ref{thm:cov}, we have, for  $r=1,2,3,4$ and $s_1,s_2,\ldots,s_r=1,\ldots,d$,
${\rm Var}\left(\sum_{i=1}^{n'}V_{i,\ell}^{(s_1)}\cdots V_{i,\ell}^{(s_r)}/n'\right)=O(\ell n^{-1})$. 
\end{lem}
Lemma~\ref{lem:lahiri} follows immediately from Lemma~3.1 of Lahiri (2003, Section~3.2.1).
\begin{lem}
\label{lem:mom}
Under the conditions of Theorem~\ref{thm:cov}, we have, for $r,s,t,u=1,\ldots,d$ and as $m\rightarrow\infty$,
\begin{eqnarray*}
\expn\left[V_{1,m}^{(r)}V_{1,m}^{(s)}\right]&=&\chi_{2,1}^{r,s}+m^{-1}\chi_{2,2}^{r,s}+O(m^{-2}),\\
\expn\left[V_{1,m}^{(r)}V_{1,m}^{(s)}V_{1,m}^{(t)}\right]&=&m^{-1/2}\chi_{3,1}^{r,s,t}+O(m^{-3/2}),\\
\expn\left[V_{1,m}^{(r)}V_{1,m}^{(s)}V_{1,m}^{(t)}V_{1,m}^{(u)}\right]&=&\chi_{4,1}^{r,s,t,u}+O(m^{-1}),
\end{eqnarray*}
where $\chi_{2,1}^{r,s}$, $\chi_{2,2}^{r,s}$ and $\chi_{3,1}^{r,s,t}$ are constants independent of $m$, and
$\chi_{4,1}^{r,s,t,u}=\chi_{2,1}^{r,s}\chi_{2,1}^{t,u}+\chi_{2,1}^{r,t}\chi_{2,1}^{s,u}+\chi_{2,1}^{r,u}\chi_{2,1}^{s,t}$.
\end{lem}
Lemma~\ref{lem:mom} can be established using arguments similar to those for proving the univariate case: see G\"{o}tze and Hipp (1983).

A generic first-level block bootstrap series ${\cal X}^*$ can be
represented as the ordered sequence of observations in $(Y_{N_1,\ell},\ldots,Y_{N_b,\ell})$, where $N_1,\ldots,N_b$ are independent random variables uniformly distributed over
$\{1,2,\ldots,n'\}$. 
Define, for  $i=0,\pm 1,\ldots\:$, $r=1,2,\ldots\:$  and $s_1,s_2,\ldots,s_r=1,\ldots,d$,
\[Q_i^{s_1,\ldots,s_r}=\sum_{j=1}^{\ell'}V_{i+j-1,k}^{(s_1)}\cdots V_{i+j-1,k}^{(s_r)}/\ell'-\expn\left[V_{1,k}^{(s_1)}\cdots V_{1,k}^{(s_r)}\right],\]
\[Q^{s_1,\ldots,s_r}=(n')^{-1}\sum_{i=1}^{n'}Q_{i}^{s_1,\ldots,s_r},\mbox{\ \ \ \ }\tilde{Q}^{s_1,\ldots,s_r}=b^{-1}\sum_{j=1}^bQ_{N_j}^{s_1,\ldots,s_r}-Q^{s_1,\ldots,s_r}\]
and $P^{s_1,\ldots,s_r}=\sum_{i=1}^{n'}V^{(s_1)}_{i,\ell}\cdots V^{(s_r)}_{i,\ell}/n'-\expn\left[V^{(s_1)}_{1,\ell}\cdots V^{(s_r)}_{1,\ell}\right]$.
Write $\breve\ell=\min(\ell',k)$ and $\bar\ell=\max(\ell',k)$.
\begin{lem}
\label{lem:size}
Under the conditions of Theorem~\ref{thm:cov}, we have
$Q^{s_1,\ldots,s_r}=O_p\left(n^{-1/2}k^{1/2}\right)$ and $\tilde{Q}^{s_1,\ldots,s_r}=O_p\left(n^{-1/2}(\ell\breve\ell/\ell')^{1/2}\right)$ 
for $r=1,2,3,4$  and $s_1,s_2,\ldots,s_r=1,\ldots,d$.
\end{lem}
{\em Proof of Lemma~\ref{lem:size}\/}.\\
For $r,s,\ldots=1,\ldots,d$ and $-\infty<i_1,i_2,\ldots<\infty$, write $\xi_{i_1,i_2,\ldots}^{r,s,\ldots}=Z_{i_1}^{(r)}Z_{i_2}^{(s)}\cdots$ and
$\bar\xi_{i_1,i_2,\ldots}^{r,s,\ldots}=\xi_{i_1,i_2,\ldots}^{r,s,\ldots}-\expn\xi_{i_1,i_2,\ldots}^{r,s,\ldots}$.
Consider first 
\begin{eqnarray*}
\expn \left(Q_1^{s_1,\ldots,s_r}\right)^2&=&k^{-r}(\ell')^{-2}\sum_{i,j=1}^{\ell'}\;\:\sum_{i_1,\ldots,i_r,j_1,\ldots,j_r=0}^{k-1}
\expn\left[\bar\xi^{s_1,\ldots,s_r}_{i+i_1,\ldots,i+i_r}\bar\xi^{s_1,\ldots,s_r}_{j+j_1,\ldots,j+j_r}\right]\\
&=&O\left\{k^{-r+1}(\ell')^{-1}\sum_{j=1}^{\ell'}\;\:\sum_{i_2,\ldots,i_r,j_1,\ldots,j_r=0}^{k-1}
\expn\left[\bar\xi^{s_1,\ldots,s_r}_{0,i_2,\ldots,i_r}\bar\xi^{s_1,\ldots,s_r}_{j+j_1,\ldots,j+j_r}\right]\right\},
\end{eqnarray*}
which follows by stationarity of the series $\{Z_j:-\infty<j<\infty\}$ and a backward shift of $i+i_1$ units. Under the assumed mixing conditions,
the last expectation has order $O(n^{-K})$ for arbitrarily large $K$ if the observations in $\bar\xi^{s_1,\ldots,s_r}_{0,i_2,\ldots,i_r}$
and $\bar\xi^{s_1,\ldots,s_r}_{j+j_1,\ldots,j+j_r}$ are at least $K\log n$ units apart.
We can therefore restrict, up to $O(n^{-K})$, the first sum to that over $j=1,\ldots,\breve\ell$, so that $\expn \left(Q_1^{s_1,\ldots,s_r}\right)^2$ has order
\begin{eqnarray}
\lefteqn{O\left\{k^{-r+1}(\ell')^{-1}\breve\ell\sum_{i_2,\ldots,i_r,j_1,\ldots,j_r=0}^{k-1}
\expn\left|\xi^{s_1,\ldots,s_r,s_1,\ldots,s_r}_{0,i_2,\ldots,i_r,j_1,\ldots,j_r}\right|\right\}}\nonumber\\
&=&O\left\{(\ell')^{-1}\breve\ell\max_{p,q\in\{s_1,\ldots,s_r\}}\left(\sum_{j=-\infty}^{\infty}\expn\left|\xi^{p,q}_{0,j}\right|\right)^r\right\}
=O\left(\breve\ell/\ell'\right).
\label{eq:qi.size}
\end{eqnarray}
Noting that 
\[{\rm Var}^*(\tilde{Q}^{s_1,\ldots,s_r})=(bn')^{-1}\sum_{i=1}^{n'}\left(Q_i^{s_1,\ldots,s_r}-Q^{s_1,\ldots,s_r}\right)^2
=O_p\left(\expn \left(Q_1^{s_1,\ldots,s_r}\right)^2/b\right),\]
we have $\tilde{Q}^{s_1,\ldots,s_r}=O_p\left(b^{-1/2}(\breve\ell/\ell')^{1/2}\right)=
O_p\left(n^{-1/2}(\ell\breve\ell/\ell')^{1/2}\right)$.

Using similar arguments, we see that
\begin{eqnarray*}
\expn \left(Q^{s_1,\ldots,s_r}\right)^2&=&O\left\{k^{-r+1}(n')^{-1}\sum_{t=0}^{\bar\ell-1}\;\:\sum_{i_2,\ldots,i_r,j_1,\ldots,j_r=0}^{k-1}
\expn\left[\bar\xi^{s_1,\ldots,s_r}_{0,i_2,\ldots,i_r}\bar\xi^{s_1,\ldots,s_r}_{t+j_1,\ldots,t+j_r}\right]
\right\}\\
&=&O\left\{k^{-r+2}(n')^{-1}\sum_{i_2,\ldots,i_r,j_1,\ldots,j_r=0}^{k-1}
\expn\left|\xi^{s_1,\ldots,s_r,s_1,\ldots,s_r}_{0,i_2,\ldots,i_r,j_1,\ldots,j_r}\right|\right\}
=O(k/n'),
\end{eqnarray*}
so that $Q^{s_1,\ldots,s_r}=O_p\left((k/n')^{1/2}\right)$.\hfill\rule{0.5ex}{2ex}

\begin{lem}
\label{lem:mom*}
Under the conditions of Theorem~\ref{thm:cov}, we have, for $r,s,t,u=1,\ldots,d$,
\begin{eqnarray*}
\expn^*\left[S^{*(r)}_nS^{*(s)}_n\right]&=&\expn\left[S^{(r)}_nS^{(s)}_n\right]+P^{r,s}+\ell^{-1}\chi_{2,2}^{r,s}+O_p(\ell n^{-1}+\ell^{-2}),\\
\expn^*\left[S^{*(r)}_nS^{*(s)}_nS^{*(t)}_n\right]&=&\expn\left[S^{(r)}_nS^{(s)}_nS^{(t)}_n\right]+O_p(\ell n^{-1}+\ell^{-1}n^{-1/2}),\\
\expn^*\left[S^{*(r)}_nS^{*(s)}_nS^{*(t)}_nS^{*(u)}_n\right]&=&\expn\left[S^{(r)}_nS^{(s)}_nS^{(t)}_nS^{(u)}_n\right]+O_p(\ell^{1/2} n^{-1/2}+\ell^{-1}).
\end{eqnarray*}
\end{lem}
{\em Proof of Lemma~\ref{lem:mom*}\/}.\\
Note first that, by Lemma~\ref{lem:lahiri}, $P^r,P^{r,s},P^{r,s,t}$ and $P^{r,s,t,u}$ have order $O_p(\ell^{1/2}n^{-1/2})$ for $r,s,t,u=1,\ldots,d$.
Lemma~\ref{lem:mom} then implies that
\begin{eqnarray*}
\expn^*\left[S^{*(r)}_nS^{*(s)}_n\right]&=&P^{r,s}+\expn\left[V^{(r)}_{1,\ell}V^{(s)}_{1,\ell}\right]-P^rP^s\\
&=&P^{r,s}+\chi_{2,1}^{r,s}+\ell^{-1}\chi_{2,2}^{r,s}+O_p(\ell n^{-1}+\ell^{-2}).
\end{eqnarray*}
By Lemma~\ref{lem:mom} again, we have $\expn\left[S^{(r)}_nS^{(s)}_n\right]=\chi_{2,1}^{r,s}+O(n^{-1})$ and the first result follows.

Similarly, the second and third results follow by noting Lemma~\ref{lem:mom} and that
\begin{eqnarray*}
\expn^*\left[S^{*(r)}_nS^{*(s)}_nS^{*(t)}_n\right]&=&b^{-1/2}\expn\left[V^{(r)}_{1,\ell}V^{(s)}_{1,\ell}V^{(t)}_{1,\ell}\right]+O_p(b^{-1/2}\ell^{1/2}n^{-1/2})\\
&=& n^{-1/2}\chi^{r,s,t}_{3,1}+O_p(\ell^{-1}n^{-1/2}+\ell n^{-1})
\end{eqnarray*}
and
\begin{eqnarray*}
\lefteqn{\expn^*\left[S^{*(r)}_nS^{*(s)}_nS^{*(t)}_nS^{*(u)}_n\right]}\\
&=& b^{-1}\left\{\chi^{r,s,t,u}_{4,1}+O(\ell^{-1})+O_p(\ell^{1/2}n^{-1/2})\right\}\\
&&\;\;+\,(1-b^{-1})\left\{\chi_{2,1}^{r,s}\chi_{2,1}^{t,u}+\chi_{2,1}^{r,t}\chi_{2,1}^{s,u}+\chi_{2,1}^{r,u}\chi_{2,1}^{s,t}+O_p(\ell^{-1}+\ell^{1/2}n^{-1/2})\right\}\\
&=&\chi^{r,s,t,u}_{4,1}+O_p(\ell^{-1}+\ell^{1/2}n^{-1/2}).
\end{eqnarray*}
\hfill\rule{0.5ex}{2ex}

A generic second-level block bootstrap series ${\cal X}^{**}$ can be
identified as the ordered sequence of observations in $(Y^*_{I_1,J_1,k},\ldots,Y^*_{I_c,J_c,k})=
(Y_{N_{I_1}+J_1,k},\ldots,Y_{N_{I_c}+J_c,k})$, where the $I_j$ and $J_j$ are independent random numbers distributed uniformly over
$\{1,2,\ldots,b\}$ and $\{1,2,\ldots,\ell'\}$ respectively, both independently of $(N_1,\ldots,N_b)$. Thus we can write
$S^{**}_n=c^{-1/2}\sum_{i=1}^cV_{N_{I_i}+J_i-1,k}-c^{1/2}(b\ell')^{-1}\sum_{i=1}^b\sum_{j=1}^{\ell'}V_{N_i+j-1,k}$.
\begin{lem}
\label{lem:mom**}
Under the conditions of Theorem~\ref{thm:cov}, we have, for $r,s,t,u=1,\ldots,d$,
\begin{eqnarray*}
\expn^{**}\left[S^{**(r)}_nS^{**(s)}_n\right]&=&\expn^*\left[S^{*(r)}_nS^{*(s)}_n\right]-P^{r,s}+\tilde{Q}^{r,s}+Q^{r,s}\\
&&\;\;+\,(k^{-1}-\ell^{-1})\chi_{2,2}^{r,s}+O_p(\ell n^{-1}+k^{-2}),\\
\expn^{**}\left[S^{**(r)}_nS^{**(s)}_nS^{**(t)}_n\right]&=&\expn^*\left[S^{*(r)}_nS^{*(s)}_nS^{*(t)}_n\right]+O_p(\ell n^{-1}+k^{-1}n^{-1/2}),\\
\expn^{**}\left[S^{**(r)}_nS^{**(s)}_nS^{**(t)}_nS^{**(u)}_n\right]&=&\expn^*\left[S^{*(r)}_nS^{*(s)}_nS^{*(t)}_nS^{*(u)}_n\right]+O_p(\ell^{1/2} n^{-1/2}+k^{-1}).
\end{eqnarray*}
\end{lem}
{\em Proof of Lemma~\ref{lem:mom**}\/}.\\
It follows from Lemmas~\ref{lem:mom} and \ref{lem:size} that
\begin{eqnarray*}
\expn^{**}\left[S^{**(r)}_nS^{**(s)}_n\right]
&=&\tilde{Q}^{r,s}+Q^{r,s}+\expn\left[V^{(r)}_{1,k}V^{(s)}_{1,k}\right]-\left(\tilde{Q}^r+Q^{r}\right)
\left(\tilde{Q}^s+Q^{s}\right)\\
&=&\tilde{Q}^{r,s}+Q^{r,s}+\expn\left[S^{(r)}_nS^{(s)}_n\right]+k^{-1}\chi_{2,2}^{r,s}\\
&&\;\;\;+\,O_p\left(k^{-2}+n^{-1}k+n^{-1}\ell\breve\ell/\ell'\right).
\end{eqnarray*}
The first result then follows by subtracting the expression for $\expn^*\left[S^{*(r)}_nS^{*(s)}_n\right]$ stated in Lemma~\ref{lem:mom*}.

Similar arguments show that
\begin{eqnarray*}
\expn^{**}\left[S^{**(r)}_nS^{**(s)}_nS^{**(t)}_n\right]&=&c^{-1/2}\left\{\expn\left[V^{(r)}_{1,k}V^{(s)}_{1,k}V^{(t)}_{1,k}\right]+O_p(\tilde{Q}^{r,s,t}+Q^{r,s,t})\right\}\\
&=& \expn\left[S^{(r)}_nS^{(s)}_nS^{(t)}_n\right]\\
&&\;\;\;+\,O_p\left(k^{-1}n^{-1/2}+kn^{-1}+n^{-1}(k\ell\breve\ell/\ell')^{1/2}\right)
\end{eqnarray*}
and
\begin{eqnarray*}
\expn^{**}\left[S^{**(r)}_nS^{**(s)}_nS^{**(t)}_nS^{**(u)}_n\right]&=&
\chi_{2,1}^{r,s}\chi_{2,1}^{t,u}+\chi_{2,1}^{r,t}\chi_{2,1}^{s,u}+\chi_{2,1}^{r,u}\chi_{2,1}^{s,t}\\
&&\;\;+\,O_p\left(k^{-1}+k^{1/2}n^{-1/2}+n^{-1/2}(\ell\breve\ell/\ell')^{1/2}\right),
\end{eqnarray*}
which, on subtracting $\expn^*\left[S^{*(r)}_nS^{*(s)}_nS^{*(t)}_n\right]$ and
$\expn^*\left[S^{*(r)}_nS^{*(s)}_nS^{*(t)}_nS^{*(u)}_n\right]$ as expressed in Lemma~\ref{lem:mom*}, yield the other two results.
\hfill\rule{0.5ex}{2ex}

Set $L_n=K\log n$ for some large $K>0$.
For $r,s,\ldots=1,\ldots,d$ and $-\infty<p,q,i,i_1,i_2,\ldots<\infty$, recall the definitions of $\xi_{i_1,i_2,\ldots}^{r,s,\ldots}$
and $\bar\xi_{i_1,i_2,\ldots}^{r,s,\ldots}$ in the proof of Lemma~\ref{lem:size}, 
and split the sum $S_n=S_{i,p,q}+\bar{S}_{i,p,q}=S_{i,p}+\bar{S}_{i,p}$ such that
\[S_{i,p,q}=n^{-1/2}\sum_{|j-(i+p)|\vee |j-(i+q)|\leq L_n}Z_j\mbox{\ \ \ and\ \ \ }S_{i,p}=n^{-1/2}\sum_{|j-(i+p)|\leq L_n}Z_j.\]
\begin{lem}
\label{lem:char}
Under the conditions of Theorem~\ref{thm:cov}, we have, for $r,s,t,u=1,\ldots,d$,
\begin{equation}
(n'\ell)^{-1}\sum_{i=1}^{n'}\sum_{p=0}^{\ell-1}Z^{(r)}_{i+p}S_{i,p}^{(s)}=n^{-1/2}\chi^{r,s}_{2,1}+O_p(\ell n^{-3/2}+n^{-1}L_n^{1/2}),
\label{eq:char11}
\end{equation}
\begin{equation}
(n'\ell)^{-1}\sum_{i=1}^{n'}\sum_{p=0}^{\ell-1}Z^{(r)}_{i+p}S_{i,p}^{(s)}S_{i,p}^{(t)}=O_p(n^{-1}),
\label{eq:char12}
\end{equation}
\begin{eqnarray}
\lefteqn{(n'\ell)^{-1}\sum_{i=1}^{n'}\sum_{p,q=0}^{\ell-1}\bar\xi^{r,s}_{i+p,i+q}S_{i,p,q}^{(t)}}\nonumber\\
&=&n^{-1/2}\sum_{i,j=-\infty}^\infty\gamma^{r,s,t}_{i,j}+O_p(\ell^{-1}n^{-1/2}L_n+\ell n^{-1}),
\label{eq:char21}
\end{eqnarray}
\begin{equation}
(n'\ell)^{-1}\sum_{i=1}^{n'}\sum_{p,q=0}^{\ell-1}\bar\xi^{r,s}_{i+p,i+q}S_{i,p,q}^{(t)}S_{i,p,q}^{(u)}=O_p(\ell n^{-1}).
\label{eq:char22}
\end{equation}
\end{lem}
{\em Proof of Lemma~\ref{lem:char}\/}.\\
We outline the proof of (\ref{eq:char21}) and (\ref{eq:char22}); that of (\ref{eq:char11}) and (\ref{eq:char12}) follows by similar, albeit simpler, arguments.

Consider first $\Pi^{r,s}_t\equiv\sum_{i=1}^{n'}\sum_{p,q=0}^{\ell-1}\sum_j^{(i+p,i+q)}\bar\xi^{r,s}_{i+p,i+q}Z^{(t)}_{j}$, where 
$\sum_j^{(i_1,i_2)}$ denotes summation over $j$ satisfying $|j-i_1|\vee|j-i_2|\leq L_n$. Note that the variance of $\Pi^{r,s}_t$ has leading term
\begin{eqnarray*}
\lefteqn{n'\ell\sum_{|q|\leq\ell}\;\;\;{\sum_j}^{(0,q)}\;\;\;
\sum_{|i'|\leq n'}\;\;\;\sum_{|p'|\vee|q'|\leq\ell}\;\;\;{\sum_{j'}}^{(i'+p',i'+q')}}\\
\lefteqn{\mbox{\ \ \ \ \ \ \ }\expn\left(\bar\xi^{r,s}_{0,q}Z^{(t)}_{j}-\expn[\bar\xi^{r,s}_{0,q}Z^{(t)}_{j}]\right)
\left(\bar\xi^{r,s}_{i'+p',i'+q'}Z^{(t)}_{j'}-\expn[\bar\xi^{r,s}_{i'+p',i'+q'}Z^{(t)}_{j'}]\right)}\nonumber\\
&\sim &
n'\ell^2\sum_{|q|\leq\ell}{\sum_j}^{(0,q)}\!\!\!\!
\sum_{|p'|\vee|q'|\leq\ell}{\sum_{j'}}^{(p',q')}\\
&&\mbox{\ \ \ \ \ \ \ }\expn\left(\bar\xi^{r,s}_{0,q}Z^{(t)}_{j}-\expn[\bar\xi^{r,s}_{0,q}Z^{(t)}_{j}]\right)
\left(\bar\xi^{r,s}_{p',q'}Z^{(t)}_{j'}-\expn[\bar\xi^{r,s}_{p',q'}Z^{(t)}_{j'}]\right)\\
&=&O\left\{n'\ell^2\sum_{|q|\leq\ell}{\sum_j}^{(0,q)}\!\!\!\!
\sum_{|p'|\vee|q'|\leq\ell}\;{\sum_{j'}}^{(p',q')}\expn\left|\xi^{r,s,t,r,s,t}_{0,q,j,p',q',j'}\right|\right\}\\
&=&O\left\{n'\ell^4\max_{u,v\in\{r,s,t\}}\left(\sum_{j=-\infty}^\infty\expn\left|\xi^{u,v}_{0,j}\right|\right)^3\right\}=O(\ell^4n),
\end{eqnarray*}
using stationarity properties and the fact that if both $i'+p'$ and $i'+q'$ differ by at least $3L_n$ from 0 and $q$, then
\[\expn\left(\bar\xi^{r,s}_{0,q}Z^{(t)}_{j}-\expn[\bar\xi^{r,s}_{0,q}Z^{(t)}_{j}]\right)
\left(\bar\xi^{r,s}_{i'+p',i'+q'}Z^{(t)}_{j'}-\expn[\bar\xi^{r,s}_{i'+p',i'+q'}Z^{(t)}_{j'}]\right)=O(n^{-K})\] for arbitrarily large $K>0$ under the assumed mixing 
conditions.
On the other hand, $\Pi^{r,s}_t$ has mean
\[ (n+O(\ell))\sum_{p=0}^{\ell-1}\sum_{q=-p}^{\ell-1-p}{\sum_j}^{(0,q)}\gamma^{r,s,t}_{q,j}
=(n+O(\ell))(\ell+O(L_n))\sum_{|q|\leq L_n}\sum_{|j-q|\leq L_n}\gamma^{r,s,t}_{q,j}.\]
It follows that $\Pi^{r,s}_t$ has expansion $n\ell\sum_{i,j=-\infty}^\infty\gamma^{r,s,t}_{i,j}+O(nL_n)+O_p(\ell^2 n^{1/2})$, which yields
(\ref{eq:char21}) on multiplying it by $n^{-1/2}(n'\ell)^{-1}$.

Consider next 
$\Pi^{r,s}_{t,u}\equiv\sum_{i=1}^{n'}\sum_{p,q=0}^{\ell-1}\sum_{j_1,j_2}^{(i+p,i+q)}\bar\xi^{r,s}_{i+p,i+q}\xi^{t,u}_{j_1,j_2}$,
which has mean of order
\[n'\ell\sum_{|q|\leq\ell}\;\;{\sum_{j_1,j_2}}^{(0,q)}\left|\expn\left[
\bar\xi^{r,s}_{0,q}\xi^{t,u}_{j_1,j_2}\right]\right|=
O\left\{n'\ell^2\sum_{j=-\infty}^\infty\expn\left|\xi^{r,t}_{0,j}\right|\sum_{i=-\infty}^\infty\expn\left|\xi^{s,u}_{0,i}\right|\right\}=O(\ell^2n),\]
and variance of order
\begin{eqnarray*}
\lefteqn{n'\ell^2\sum_{|q|\leq\ell}\;{\sum_{j_1,j_2}}^{(0,q)}\!\!\!\!\!
\sum_{|p'|\vee|q'|\leq\ell}\;{\sum_{j'_1,j'_2}}^{(p',q')}}\\
\lefteqn{\mbox{\ \ \ \ \ \ \ \ \ \ }\expn\left(\bar\xi^{r,s}_{0,q}\xi^{t,u}_{j_1,j_2}-\expn[\bar\xi^{r,s}_{0,q}\xi^{t,u}_{j_1,j_2}]\right)
\left(\bar\xi^{r,s}_{p',q'}\xi^{t,u}_{j'_1,j'_2}-\expn[\bar\xi^{r,s}_{p',q'}\xi^{t,u}_{j'_1,j'_2}]\right)}\\
&=&O\left\{n'\ell^2\!\!\!\!\sum_{|q|,|p'|,|q'|\leq\ell}\;{\sum_{j_1,j_2}}^{(0,q)}
\;{\sum_{j'_1,j'_2}}^{(p',q')}\expn\left|\xi^{r,s,t,u,r,s,t,u}_{0,q,j_1,j_2,p',q',j'_1,j'_2}\right|\right\}=O(\ell^5n).
\end{eqnarray*}
Thus (\ref{eq:char22}) follows by multiplying $\Pi^{r,s}_{t,u}$ by $(n'\ell)^{-1}n^{-1}$.
\hfill\rule{0.5ex}{2ex}

Consider next the decomposition
$S_n={\cal S}_{i,j,p,q}+\bar{\cal S}_{i,j,p,q}={\cal S}_{i,j,p}+\bar{\cal S}_{i,j,p}$ such that
${\cal S}_{i,j,p,q}=n^{-1/2}\sum_t^{(i+j-1+p,i+j-1+q)}Z_t$ and ${\cal S}_{i,j,p}=n^{-1/2}\sum_{|t-(i+j-1+p)|\leq L_n}Z_t$,
for $-\infty<p,q,i,j<\infty$. Arguments similar to those for proving Lemma~\ref{lem:char} can be used to establish:
\begin{lem}
\label{lem:char*}
Under the conditions of Theorem~\ref{thm:cov}, we have, for $r,s,t,u=1,\ldots,d$,
\[
(n'\ell'k)^{-1}\sum_{i=1}^{n'}\sum_{j=1}^{\ell'}\sum_{p=0}^{k-1}Z^{(r)}_{i+j-1+p}{\cal S}_{i,j,p}^{(s)}=n^{-1/2}\chi^{r,s}_{2,1}+O_p(\ell  n^{-3/2}+n^{-1}L_n^{1/2}),
\]
\[
(n'\ell' k)^{-1}\sum_{i=1}^{n'}\sum_{j=1}^{\ell'}\sum_{p=0}^{k-1}Z^{(r)}_{i+j-1+p}{\cal S}_{i,j,p}^{(s)}{\cal S}_{i,j,p}^{(t)}=O_p(n^{-1}),
\]
\begin{eqnarray*}
\lefteqn{(n'\ell'k)^{-1}\sum_{i=1}^{n'}\sum_{j=1}^{\ell'}\sum_{p,q=0}^{k-1}\bar\xi^{r,s}_{i+j-1+p,i+j-1+q}{\cal S}_{i,j,p,q}^{(t)}}\nonumber\\
&=&n^{-1/2}\sum_{i,j=-\infty}^\infty\gamma^{r,s,t}_{i,j}+O_p(k^{-1}n^{-1/2}L_n+k n^{-1}+\ell n^{-3/2}),
\end{eqnarray*}
\[
(n'\ell' k)^{-1}\sum_{i=1}^{n'}\sum_{j=1}^{\ell'}\sum_{p,q=0}^{k-1}\bar\xi^{r,s}_{i+j-1+p,i+j-1+q}{\cal S}_{i,j,p,q}^{(t)}{\cal S}_{i,j,p,q}^{(u)}=O_p(k n^{-1}).
\]
\end{lem}

We now proceed with the proof of Theorem~\ref{thm:cov}.

Define $H_r(x)=(\partial/\partial x^{(r)})H(x)$, $H_{rs}=(\partial^2/\partial x^{(r)}\partial x^{(s)})H(x)$,
etc., for $r,s,\ldots=1,\ldots,d$.  Recall that we write $H_r=H_r(\mu)$, $H_{rs}=H_{rs}(\mu)$, etc.\ for convenience. Note that $\hat\mu^{(r)}\equiv\expn^*\bar{X}^{*(r)}=
\ell^{-1/2}P^r+\mu^{(r)}$. Write $\hat{H}_r=H_r(\hat\mu)$, $\hat{H}_{rs}=H_{rs}(\hat\mu)$, etc. Taylor expansion shows that ${\rm Var}(n^{1/2}\hat\theta)$ has leading term
$\sigma^2=\sum_{r,s=1}^dH_rH_s\expn\left[S^{(r)}_nS^{(s)}_n\right]$. Define $\hat\sigma^2=\sum_{r,s=1}^d\hat{H}_r\hat{H}_s\expn^*\left[S^{*(r)}_nS^{*(s)}_n\right]$,
which can, by Lemmas~\ref{lem:mom} and \ref{lem:mom*}, be Taylor expanded to give
\begin{equation}
\label{eq:sigmahat}
 \hat\sigma^2=\sigma^2+\sum_{r,s=1}^dH_rH_s\left(\ell^{-1}\chi^{r,s}_{2,2}+P^{r,s}\right)+2\ell^{-1/2}\!\!\!\sum_{r,s,t=1}^dH_rH_{st}\chi^{r,s}_{2,1}P^t
+O_p(\ell n^{-1}+\ell^{-2}). 
\end{equation}
Lahiri (2003, Section~6.4.3) provides an Edgeworth expansion for the distribution function $G$ of $n^{1/2}(\hat\theta-\theta)$:
\begin{equation}
G(x)=\Phi(x/\sigma)-n^{-1/2}\left[{\cal K}_{31}+{\cal K}_{32}(x^2/\sigma^2-1)\right]\phi(x/\sigma)+O(n^{-1}),
\label{eq:edge}
\end{equation}
where ${\cal K}_{31}$ and ${\cal K}_{32}$  are smooth functions, both of order $O(1)$, of 
the moments 
$\expn\left[S^{(s_1)}_n\cdots S^{(s_r)}_n\right]$, for $s_1,\ldots,s_r=1,\ldots,d$ and $r=2,3,4$,
and $\Phi$ denotes the standard normal distribution function.
Lahiri's (2003) Theorem~6.7 derives a block bootstrap version of (\ref{eq:edge}) under the conditions of our Theorem~\ref{thm:cov}:
\begin{equation}
G^*(x)=\Phi(x/\hat\sigma)-n^{-1/2}\left[\hat{\cal K}_{31}+\hat{\cal K}_{32}(x^2/\hat\sigma^2-1)\right]\phi(x/\hat\sigma)+O_p(\ell n^{-1}),
\label{eq:edge*}
\end{equation}
where $\hat{\cal K}_{31}$ and $\hat{\cal K}_{32}$ have the same expressions as ${\cal K}_{31}$ and ${\cal K}_{32}$ with the population moments
$\expn\left[S^{(s_1)}_n\cdots S^{(s_r)}_n\right]$ replaced by $\expn^*\left[S^{*(s_1)}_n\cdots S^{*(s_r)}_n\right]$.
With the aid of Lemma~\ref{lem:mom*} and the expressions (\ref{eq:sigmahat}),  (\ref{eq:edge}) and (\ref{eq:edge*}), we can expand the difference between $G^{*-1}$ and $G^{-1}$,
so that, for $\xi\in(0,1)$,
\begin{equation}
\label{eq:cdf.sn}
\prob\left(n^{1/2}(\hat\theta-\theta)\leq G^{*-1}(\xi)\right)=\prob(T_n\leq y)+O(\ell n^{-1}+\ell^{-2}), 
\end{equation}
where $T_n=n^{1/2}(\hat\theta-\theta)-z_\xi(2\sigma)^{-1}\left(\sum_{r,s=1}^dH_rH_sP^{r,s}+2\ell^{-1/2}\!\sum_{r,s,t=1}^dH_rH_{st}\chi^{r,s}_{2,1}P^t\right)$ and
$y=G^{-1}(\xi)+\ell^{-1}z_\xi(2\sigma)^{-1}\sum_{r,s=1}^dH_rH_s\chi^{r,s}_{2,2}$.
Noting that $P^r$ and $P^{r,s}$ are $O_p(\ell^{1/2}n^{-1/2})$ by Lemma~\ref{lem:size}, that $n^{1/2}(\hat\theta-\theta)=\sum_{u=1}^dH_uS_n^{(u)}+O_p(n^{-1/2})$ and 
expanding the characteristic function of $T_n$ about that of $n^{1/2}(\hat\theta-\theta)$, we get, for $\beta\in{\Bbb R}$,
\begin{eqnarray}
\lefteqn{\expn\, e^{\iota \beta T_n}-\expn\, e^{\iota\beta n^{1/2}(\hat\theta-\theta)}}\nonumber\\
&=&-\,\iota \beta z_\xi(2\sigma)^{-1}\sum_{r,s=1}^dH_rH_s\expn\!\left[P^{r,s}
\exp\left(\iota \beta \sum_{u=1}^d H_uS_n^{(u)}\right)\right]\nonumber\\
&&-\,\iota \beta z_\xi\sigma^{-1}\ell^{-1/2}\!\!\!\sum_{r,s,t=1}^d\!\!H_rH_{st}\chi^{r,s}_{2,1}\expn\!\left[P^t\exp\left(\iota \beta \sum_{u=1}^d H_uS_n^{(u)}\right)\right]\nonumber\\
&&+\,O(\ell n^{-1}).
\label{eq:charfn}
\end{eqnarray}
Note that for $s_1,s_2,s_3=1,\ldots,d$,
\begin{eqnarray}
\lefteqn{(n'\ell)^{-1}\left|\sum_{i=1}^{n'}\sum_{p,q=0}^{\ell-1}\expn\left[\bar{\xi}^{r,s}_{i+p,i+q}S_{i,p,q}^{(s_1)}S_{i,p,q}^{(s_2)} S_{i,p,q}^{(s_3)}\right]\right|}\nonumber\\
&\leq &(n'\ell)^{-1}n^{-3/2}\sum_{i=1}^{n'}\sum_{p,q=0}^{\ell-1}\;\;{\sum_{i_1,i_2,i_3}}^{(i+p,i+q)}\expn\left|\bar{\xi}^{r,s}_{i+p,i+q}\xi^{s_1,s_2,s_3}_{i_1,i_2,i_3}\right|
\nonumber\\
&=&O\left\{n^{-3/2}\sum_{|q|\leq\ell}\;\;{\sum_{i_1,i_2,i_3}}^{(0,q)}\expn\left|\bar{\xi}^{r,s}_{0,q}\xi^{s_1,s_2,s_3}_{i_1,i_2,i_3}\right|\right\}
=O(\ell n^{-3/2}L_n^3).
\label{eq:remainder}
\end{eqnarray}
It follows by expansion of the exponential function, (\ref{eq:char21}), (\ref{eq:char22}) and (\ref{eq:remainder}) that
\begin{eqnarray}
\lefteqn{\expn\!\left[P^{r,s}
\exp\left(\iota \beta \sum_{u=1}^d H_uS_n^{(u)}\right)\right]}\nonumber\\
&=&(n'\ell)^{-1}\sum_{i=1}^{n'}\sum_{p,q=0}^{\ell-1} \expn\left[\bar{\xi}^{r,s}_{i+p,i+q}\exp\left(\iota \beta \sum_{u=1}^d H_uS_n^{(u)}\right)\right.\nonumber\\
&&\left.\left\{\iota \beta \sum_{u=1}^d H_uS_{i,p,q}^{(u)}
+2^{-1}\beta^2\sum_{t,u=1}^d H_t H_uS^{(t)}_{i,p,q}S_{i,p,q}^{(u)}+\exp\left(-\iota \beta \sum_{u=1}^d H_uS_{i,p,q}^{(u)}\right)\right\}
\right]\nonumber\\
&&\;\;\;+\,O(\ell n^{-3/2}L_n^3)\nonumber\\
&=& (n'\ell)^{-1}\sum_{i=1}^{n'}\sum_{p,q=0}^{\ell-1} \expn\left[\bar{\xi}^{r,s}_{i+p,i+q}\exp\left(\iota \beta \sum_{u=1}^d H_u\bar{S}_{i,p,q}^{(u)}\right)\right] \nonumber\\
&&+\,\iota \beta n^{-1/2}\sum_{t=1}^d H_t\!\!\sum_{i,j=-\infty}^\infty\!\!\gamma^{r,s,t}_{i,j}\,
\expn\left[e^{\iota \beta n^{1/2}(\hat\theta-\theta)}\right] +O(\ell^{-1}n^{-1/2}L_n+\ell n^{-1})\nonumber\\
&=&\iota \beta n^{-1/2}\sum_{t=1}^d H_t\!\!\sum_{i,j=-\infty}^\infty\!\!\gamma^{r,s,t}_{i,j}\,
\expn\left[e^{\iota \beta n^{1/2}(\hat\theta-\theta)}\right]+O(\ell^{-1}n^{-1/2}L_n+\ell n^{-1}).
\label{eq:charterm1}
\end{eqnarray}
The last equality follows by the assumed mixing properties and noting that observations defining $\bar{S}_{i,p,q}^{(u)}$ and $\bar{\xi}^{r,s}_{i+p,i+q}$ are at least $L_n$ units apart on the series
and that $\expn\,\bar{\xi}^{r,s}_{i+p,i+q}=0$.
Noting that
$(n'\ell)^{-1}\left|\sum_{i=1}^{n'}\sum_{p=0}^{\ell-1}\expn\left[Z^{(r)}_{i+p}S_{i,p}^{(s_1)}S_{i,p}^{(s_2)} S_{i,p}^{(s_3)}\right]\right|=O(n^{-3/2}L_n^3)$ for $s_1,s_2,s_3=1,\ldots,d$,
the same arguments show that
\begin{eqnarray}
\lefteqn{\ell^{-1/2}\expn\!\left[P^t\exp\left(\iota \beta \sum_{u=1}^d H_uS_n^{(u)}\right)\right]}\nonumber\\
&=&\iota \beta n^{-1/2}\sum_{u=1}^d H_u\chi^{t,u}_{2,1}\,
\expn\left[e^{\iota \beta n^{1/2}(\hat\theta-\theta)}\right]+O(\ell n^{-3/2}+n^{-1}L_n^{1/2}).
\label{eq:charterm2}
\end{eqnarray}
Substitution of (\ref{eq:charterm1}) and (\ref{eq:charterm2}) into (\ref{eq:charfn}) gives
\begin{eqnarray}
\expn\, e^{\iota \beta T_n}/\expn\, e^{\iota\beta n^{1/2}(\hat\theta-\theta)}&=& 1+n^{-1/2}(2\sigma)^{-1}\beta^2 z_\xi\sum_{r,s,t=1}^d
\!\!H_rH_sH_t\sum_{i,j=-\infty}^\infty\gamma^{r,s,t}_{i,j}
\nonumber\\
&&+\,n^{-1/2}\sigma^{-1} \beta^2 z_\xi\sum_{r,s,t,u=1}^d\!\!H_rH_uH_{st}\chi^{r,s}_{2,1}\chi^{t,u}_{2,1}\nonumber\\
&&+\,O(\ell n^{-1}+\ell^{-1}n^{-1/2}L_n).
\label{eq:charfn2}
\end{eqnarray}
It follows by inverse Fourier-transforming $\expn\, e^{\iota \beta T_n}$ that
\begin{eqnarray}
\prob(T_n\leq x)&=&G(x)+n^{-1/2}2^{-1}\sigma^{-4} z_\xi x\phi(x/\sigma)\times\nonumber\\
&&\left\{\sum_{r,s,t=1}^d
\!\!H_rH_sH_t\sum_{i,j=-\infty}^\infty\gamma^{r,s,t}_{i,j}+2\!\!\!\sum_{r,s,t,u=1}^d\!\!H_rH_uH_{st}\chi^{r,s}_{2,1}\chi^{t,u}_{2,1}\right\}\nonumber\\
&&+\,O(\ell n^{-1}+\ell^{-1}n^{-1/2}L_n).
\label{eq:cdf.tn}
\end{eqnarray}
It then follows by combining (\ref{eq:cdf.sn}) and (\ref{eq:cdf.tn}), setting $x=y$ and noting that $y=\sigma z_\xi+O(n^{-1/2}+\ell^{-1})$ that
\begin{eqnarray*}
\lefteqn{\prob\left(n^{1/2}(\hat\theta-\theta)\leq G^{*-1}(\xi)\right)}\nonumber\\
&=& \xi +\ell^{-1}2^{-1} \sigma^{-2}z_\xi\phi(z_\xi) \sum_{r,s=1}^dH_rH_s\chi^{r,s}_{2,2}+n^{-1/2}2^{-1}\sigma^{-3} z_\xi^2 \phi(z_\xi)\nonumber\\
&&\times\left\{\sum_{r,s,t=1}^d
\!\!H_rH_sH_t\sum_{i,j=-\infty}^\infty\gamma^{r,s,t}_{i,j}+2\!\!\!\sum_{r,s,t,u=1}^d\!\!H_rH_uH_{st}\chi^{r,s}_{2,1}\chi^{t,u}_{2,1}\right\}\nonumber\\
&&+\,O(\ell n^{-1}+\ell^{-2}),
\end{eqnarray*}
which yields (\ref{eq:boot}) on setting $\xi=1-\alpha$ and taking complement.

For proving (\ref{eq:it.boot}), write $\mu^*=\expn^{**}\bar{X}^{**}$, $H^*_r=H_r(\mu^*)$, $H^*_{rs}=H_{rs}(\mu^*)$ etc.\ and
define $\sigma^{*2}=\sum_{r,s=1}^dH^*_rH^*_s\expn^{**}\!\!\left[S^{**(r)}_nS^{**(s)}_n\right]$.
Note that, for $r=1,\ldots,d$, $\mu^{*(r)}-\hat\mu^{(r)}=k^{-1/2}\left(\tilde{Q}^{r}+Q^r\right)-\ell^{-1/2}P^r=O_p(n^{-1/2})$
by Lemmas~\ref{lem:lahiri} and \ref{lem:size}. It follows by Lemma~\ref{lem:mom**} and Taylor expansion  that
\begin{eqnarray}
\sigma^{*2}&=&\hat\sigma^2+\sum_{r,s=1}^d{H}_r{H}_s\left[\tilde{Q}^{r,s}+Q^{r,s}-P^{r,s}+(k^{-1}-\ell^{-1})\chi^{r,s}_{2,2}\right]\nonumber\\
&&+\,2\sum_{r,s,t=1}^d{H}_r{H}_{st}\chi^{r,s}_{2,1}\left[k^{-1/2}\left(\tilde{Q}^{t}+Q^t\right)-\ell^{-1/2}P^t\right]\nonumber\\
&&+\,O_p(n^{-1}\ell+k^{-2}),
\label{eq:sigma*}
\end{eqnarray}
using the fact that $\hat\mu=\mu+O_p(n^{-1/2})$.
Denote by ${\cal K}^*_{31}$ and ${\cal K}^*_{32}$ the versions of ${\cal K}_{31}$ and ${\cal K}_{32}$ with the moments 
$\expn\left[S^{(s_1)}_n\cdots S^{(s_r)}_n\right]$ replaced by  $\expn^{**}\!\!\left[S^{**(s_1)}_n\cdots S^{**(s_r)}_n\right]$ in their definitions.
Thus, by analogy with (\ref{eq:edge*}), we have
\begin{equation}
G^{**}(x)=\Phi(x/\sigma^*)-n^{-1/2}\left[{\cal K}^*_{31}+{\cal K}^*_{32}(x^2/\sigma^{*2}-1)\right]\phi(x/\sigma^*)+O_p(k n^{-1}).
\label{eq:edge**}
\end{equation}
The expansions (\ref{eq:edge*}), (\ref{eq:sigma*}), (\ref{eq:edge**}) and the results in Lemma~\ref{lem:mom**} enable us to expand $G^{**-1}(\xi)$ about
$G^{*-1}(\xi)$ and write
\begin{eqnarray}
\lefteqn{\prob^*\left((b\ell)^{1/2}[H(\bar{X}^*)-H(\expn^*\bar{X}^*)]\leq G^{**-1}(\xi)\right)}\nonumber\\
&=&\prob^*\left((b\ell)^{1/2}[H(\bar{X}^*)-H(\expn^*\bar{X}^*)]-\Delta^*_n\leq\hat{y}\right)+O_p(\ell n^{-1}+k^{-2}),
\label{eq:calib}
\end{eqnarray}
where $\Delta^*_n=b^{-1}\sum_{j=1}^bR^*_j$, 
\[R^*_j=(2\hat\sigma)^{-1}z_\xi
\left\{\sum_{r,s=1}^d{H}_r{H}_s({Q}^{r,s}_{N_j}-Q^{r,s})+2k^{-1/2}\sum_{r,s,t=1}^d{H}_r{H}_{st}\chi^{r,s}_{2,1}({Q}^{t}_{N_j}-Q^t)\right\}\]
and
\begin{eqnarray*}
\hat{y}&=&G^{*-1}(\xi)+(2\hat\sigma)^{-1}z_\xi\left\{
\sum_{r,s=1}^d{H}_r{H}_s\left[Q^{r,s}-P^{r,s}+(k^{-1}-\ell^{-1})\chi^{r,s}_{2,2}\right]\right.\\
&&\mbox{\ \ \ \ \ \ \ \ \ \ \ \ \ \ \ \ \ \ \ \ \ \ \ \ \ \ \ \ \ \ \ \ \ \ \ }\left.+\,2\sum_{r,s,t=1}^d{H}_r{H}_{st}\chi^{r,s}_{2,1}\left(k^{-1/2}Q^t-\ell^{-1/2}P^t\right)
\right\}.
\end{eqnarray*}
Define also $Y^*_j=\sum_{r=1}^d\left(V^{(r)}_{N_j,\ell}-P^r\right)\hat{H}_r$ for $j=1,\ldots,b$, so that $(b\ell)^{1/2}[H(\bar{X}^*)-H(\expn^*\bar{X}^*)]
=b^{-1/2}\sum_{j=1}^bY^*_j+O_p(n^{-1/2})$ by Taylor expansion.
Note that the observations
$(Y^*_j,R^*_j)$ are independent, zero-mean and identically distributed with respect to first-level block bootstrap sampling, conditional on ${\cal X}$.
We see by Lemma~\ref{lem:size} that $\Delta^*_n=O_p(k^{1/2}n^{-1/2})$ and by (\ref{eq:qi.size}) that $R^*_j=O_p((\breve\ell/\ell')^{1/2})$,
whereas $Y^*_j=O_p(1)$ by Lemmas~\ref{lem:lahiri} and \ref{lem:mom}.
It follows that, conditional on ${\cal X}$, 
$(b\ell)^{1/2}[H(\bar{X}^*)-H(\expn^*\bar{X}^*)]-\Delta^*_n$ and $(b\ell)^{1/2}[H(\bar{X}^*)-H(\expn^*\bar{X}^*)]$ have identical means, 
variances differing by $-2b^{-1/2}\expn^*[Y^*_1R^*_1]+O_p(kn^{-1})$ and third cumulants differing by
$-3b^{-1}\expn^*[Y^{*2}_1R^*_1]+O_p(kn^{-1})=O_p(\ell n^{-1})$.
Such cumulant differences can be employed to establish an Edgeworth expansion
for $(b\ell)^{1/2}[H(\bar{X}^*)-H(\expn^*\bar{X}^*)]-\Delta^*_n$ analogous to (\ref{eq:edge*}), bearing in mind that $\hat{\cal K}_{31}$  and $\hat{\cal K}_{32}$ stem from
the first and third cumulants respectively:
\begin{eqnarray}
\lefteqn{\prob^*\left((b\ell)^{1/2}[H(\bar{X}^*)-H(\expn^*\bar{X}^*)]-\Delta^*_n\leq x\right)}\nonumber\\
&=& G^*(x)+b^{-1/2}\hat\sigma^{-3}x\phi(x/\hat\sigma)\expn^*[Y^*_1R^*_1]+O_p(\ell n^{-1}).
\label{eq:calib2}
\end{eqnarray}
Note by Lemmas~\ref{lem:lahiri}, \ref{lem:mom} and \ref{lem:size} that for $r,s,t=1,\ldots,d$,
\begin{eqnarray}
\lefteqn{{\rm Cov}^*\left(V^{(r)}_{N_1,\ell},\,{Q}^{s,t}_{N_1}\right)}\nonumber\\
&=&(n'\ell')^{-1}\sum_{i=1}^{n'}\sum_{j=1}^{\ell'}V^{(r)}_{i,\ell}V^{(s)}_{i+j-1,k}V^{(t)}_{i+j-1,k}-P^r\expn[V^{(s)}_{1,k}V^{(t)}_{1,k}]
-P^rQ^{s,t}\nonumber\\
&=& (n'\ell'k)^{-1}\ell^{-1/2}\sum_{i=1}^{n'}\sum_{j=1}^{\ell'}\sum_{a=0}^{\ell-1}\sum_{p,q=0}^{k-1} \xi^{r,s,t}_{i+a,i+j-1+p,i+j-1+q}\nonumber\\
&&+\,O_p(\ell^{1/2}n^{-1/2}).
\label{eq:yr1}
\end{eqnarray}
Consider 
\begin{eqnarray*}
\lefteqn{\sum_{j=1}^{\ell'}\sum_{a=0}^{\ell-1}\sum_{p,q=0}^{k-1}\expn\, \xi^{r,s,t}_{a,j-1+p,j-1+q}}\\
&=& \sum_{p,q=0}^{k-1}\sum_{a=1-\ell'}^{\ell-1}\left\{(\ell-a)\wedge\ell'-(1-a)\vee 1+1\right\}\expn\, \xi^{r,s,t}_{a,p,q}\\
&=&\sum_{|p|,|q|\leq L_n}\left[\sum_{a=O(L_n)}^{k+O(L_n)}\left\{(\ell-a)\wedge\ell'-(1-a)\vee 1+1\right\}\right] \gamma^{r,s,t}_{p,q}\\
&=& \left\{k\ell'+O(\ell'L_n+L_n^2)\right\}\sum_{p,q=-\infty}^\infty \gamma^{r,s,t}_{p,q}
\end{eqnarray*}
and
\begin{eqnarray*}
\lefteqn{{\rm Var}\left( (n')^{-1}\sum_{i=1}^{n'}\sum_{j=1}^{\ell'}\sum_{a=0}^{\ell-1}\sum_{p,q=0}^{k-1} \xi^{r,s,t}_{i+a,i+j-1+p,i+j-1+q} \right)}\\
&\sim & (n')^{-1}k\ell'\sum_{|i'|\leq n'}\;\;\sum_{|j'|\leq\ell'}\;\;\sum_{|a|,|a'|\leq\ell}\;\;\sum_{|q|,|p'|,|q'|\leq k}\expn\left[\bar\xi^{r,s,t}_{a,0,q}\bar\xi^{r,s,t}_{i'+a',i'+j'-1+p',i'+j'-1+q'}\right]\\
&=&O\left\{n^{-1}k\ell'\ell\sum_{|j'|\leq\ell'}\;\;\sum_{|i'|,|a|\leq\ell}\;\;\sum_{|q|,|p'|,|q'|\leq k}\expn\left|\xi^{r,s,t,r,s,t}_{a,0,q,i',i'+j'-1+p',i'+j'-1+q'}\right|\right\}\\
&=&O\left\{n^{-1}(k\ell'\ell)^2\sum_{|q|\leq k}\expn\left|\xi^{s,t}_{0,q}\right|\sum_{|a|\leq\ell}\expn\left|  \xi^{r,r}_{a,0}\right|\sum_{|q'|\leq k}  \expn\left|\xi^{s,t}_{0,q'}\right|\right\}
=O(n^{-1}(k\ell'\ell)^2),
\end{eqnarray*}
so that
\begin{eqnarray}
\lefteqn{(n')^{-1}\sum_{i=1}^{n'}\sum_{j=1}^{\ell'}\sum_{a=0}^{\ell-1}\sum_{p,q=0}^{k-1} \xi^{r,s,t}_{i+a,i+j-1+p,i+j-1+q}}\nonumber\\
&=&k\ell'\sum_{i,j=-\infty}^\infty \gamma^{r,s,t}_{i,j}+O_p(\ell'L_n+L_n^2+k\ell'\ell n^{-1/2}).
\label{eq:yr2}
\end{eqnarray}
Similar arguments show that
\begin{eqnarray}
{\rm Cov}^*\left(V^{(r)}_{N_1,\ell},\,{Q}^{t}_{N_1}\right)
&=&(k/\ell)^{1/2}\sum_{a=-\infty}^\infty\gamma^{r,t}_a+O_p\left\{(k\ell)^{-1/2}L_n\right.\nonumber\\
&&\left.\mbox{\ \ \ \ }+\,(\ell')^{-1}(k\ell)^{-1/2}L_n^2+k^{1/2}n^{-1/2}\right\}.
\label{eq:yr3}
\end{eqnarray}
Combining (\ref{eq:yr1})--(\ref{eq:yr3}), we have
\begin{eqnarray}
\expn^*[Y^{*}_1R^*_1]&=&\ell^{-1/2}(2\hat\sigma)^{-1}z_\xi\left\{\sum_{r,s,t=1}^d{H}_r
{H}_s{H}_t\sum_{i,j=-\infty}^\infty \gamma^{r,s,t}_{i,j}\right.\nonumber\\
&&\mbox{\ \ \ \ \ \ \ \ \ \ \ \ \ \ \ \ \ \ \ \ \ \ \ \ }\left.
+\,2\sum_{r,s,t,u=1}^d{H}_r
{H}_s{H}_{tu}\chi^{r,t}_{2,1}\chi^{s,u}_{2,1}
\right\}\nonumber\\
&&+\,O_p( \ell^{-1/2}k^{-1}L_n+(k\ell')^{-1}\ell^{-1/2}L_n^2+\ell^{1/2}n^{-1/2}).
\label{eq:yr5}
\end{eqnarray}
Substitution of (\ref{eq:yr5}) into (\ref{eq:calib2}), setting $x=\hat{y}$ and noting (\ref{eq:calib}), we have
\begin{eqnarray*}
\lefteqn{\prob^*\left((b\ell)^{1/2}[H(\bar{X}^*)-H(\expn^*\bar{X}^*)]\leq G^{**-1}(\xi)\right)}\nonumber\\
&=& \xi+2^{-1}\sigma^{-2}z_\xi\phi(z_\xi)\left\{
\sum_{r,s=1}^d{H}_r{H}_s\left[Q^{r,s}-P^{r,s}+(k^{-1}-\ell^{-1})\chi^{r,s}_{2,2}\right]\right.\nonumber\\
&&\mbox{\ \ \ \ \ \ \ \ \ \ \ \ \ \ \ \ \ \ \ \ \ \ \ }\left.+\,2\sum_{r,s,t=1}^d{H}_r{H}_{st}\chi^{r,s}_{2,1}\left(k^{-1/2}Q^t-\ell^{-1/2}P^t\right)
\right\}\nonumber\\
&&+\,n^{-1/2}2^{-1}\sigma^{-3} z_\xi^2\phi(z_\xi)\left\{\sum_{r,s,t=1}^d{H}_r
{H}_s{H}_t\sum_{i,j=-\infty}^\infty \gamma^{r,s,t}_{i,j}\right.\nonumber\\
&&\mbox{\ \ \ \ \ \ \ \ \ \ \ \ \ \ \ \ \ \ \ \ \ \ \ \ \ \ \ \ \ \ \ }\left.
+\,2\sum_{r,s,t,u=1}^d{H}_r
{H}_s{H}_{tu}\chi^{r,t}_{2,1}\chi^{s,u}_{2,1}
\right\}\nonumber\\
&&+\,O_p(\ell n^{-1}+k^{-2}),
\end{eqnarray*}
inversion of which gives $\hat\alpha=\alpha+\delta_n+B_n+O_p(\ell n^{-1}+k^{-2})$, where
\begin{eqnarray*}
\delta_n&=&-(k^{-1}-\ell^{-1})2^{-1}\sigma^{-2}z_\alpha\phi(z_\alpha)
\sum_{r,s=1}^d{H}_r{H}_s\chi^{r,s}_{2,2}
+ n^{-1/2}2^{-1}\sigma^{-3} z_\alpha^2\phi(z_\alpha)\\
&&\times\left(\sum_{r,s,t=1}^d{H}_r
{H}_s{H}_t\sum_{i,j=-\infty}^\infty \gamma^{r,s,t}_{i,j}+2\sum_{r,s,t,u=1}^d{H}_r
{H}_s{H}_{tu}\chi^{r,t}_{2,1}\chi^{s,u}_{2,1}
\right),
\end{eqnarray*}
\begin{eqnarray*}
B_n&=&
-2^{-1}\sigma^{-2}z_\alpha\phi(z_\alpha)\\
&&\times\left\{
\sum_{r,s=1}^d{H}_r{H}_s\left(Q^{r,s}-P^{r,s}\right)+2\sum_{r,s,t=1}^d{H}_r{H}_{st}\chi^{r,s}_{2,1}\left(k^{-1/2}Q^t-\ell^{-1/2}P^t\right)\right\}.
\end{eqnarray*}
It follows from (\ref{eq:cdf.sn}) that the coverage probability of ${\cal I}_C(\alpha)$ is
\begin{equation}
1-\prob\left(n^{1/2}(\hat\theta-\theta)\leq G^{*-1}(1-\hat\alpha)\right)
= 1-\prob(\tilde{T}_n\leq\tilde{y})
+O(\ell n^{-1}+k^{-2}),
\label{eq:cov*}
\end{equation}
where
\begin{eqnarray*}
\tilde{T}_n&=&n^{1/2}(\hat\theta-\theta)+z_\alpha(2\sigma)^{-1}\left(\sum_{r,s=1}^dH_rH_sP^{r,s}+2\ell^{-1/2}\!\sum_{r,s,t=1}^dH_rH_{st}\chi^{r,s}_{2,1}P^t\right)\\
&&\;\;\;+\,\sigma\phi(z_\alpha)^{-1}B_n\\
&=&n^{1/2}(\hat\theta-\theta)+z_\alpha(2\sigma)^{-1}\times\\
&&\left\{\sum_{r,s=1}^dH_rH_s(2P^{r,s}-Q^{r,s})+2\sum_{r,s,t=1}^dH_rH_{st}\chi^{r,s}_{2,1}(2\ell^{-1/2}P^t-k^{-1/2}Q^t)\right\}
\end{eqnarray*}
and
$\tilde{y}=G^{-1}(1-\alpha-\delta_n)-\ell^{-1}z_\alpha(2\sigma)^{-1}\sum_{r,s=1}^dH_rH_s\chi^{r,s}_{2,2}$.
Similar to (\ref{eq:charterm1}) and (\ref{eq:charterm2}), 
$\expn\!\left[Q^{r,s}
\exp\left(\iota \beta \sum_{u=1}^d H_uS_n^{(u)}\right)\right]$ and $k^{-1/2}\expn\!\left[Q^{t}
\exp\left(\iota \beta \sum_{u=1}^d H_uS_n^{(u)}\right)\right]$ can be expanded by invoking Lemma~\ref{lem:char*}, so that the difference between the characteristic functions of $\tilde{T}_n$ and $n^{1/2}(\hat\theta-\theta)$ can be established as in the proof of (\ref{eq:charfn2}).
This enables us to derive an Edgeworth expansion for $\tilde{T}_n$ analogous to (\ref{eq:cdf.tn}):
\begin{eqnarray}
\prob(\tilde T_n\leq x)&=&G(x)-n^{-1/2}2^{-1}\sigma^{-4} z_\alpha x\phi(x/\sigma)\times\nonumber\\
&&\left\{\sum_{r,s,t=1}^d
\!\!H_rH_sH_t\sum_{i,j=-\infty}^\infty\gamma^{r,s,t}_{i,j}+2\!\!\!\sum_{r,s,t,u=1}^d\!\!H_rH_uH_{st}\chi^{r,s}_{2,1}\chi^{t,u}_{2,1}\right\}\nonumber\\
&&+O(\ell n^{-1}+k^{-1}n^{-1/2}L_n).
\label{eq:cdf.tilde.tn}
\end{eqnarray}
The coverage expansion (\ref{eq:it.boot}) for ${\cal I}_C(\alpha)$ then follows by noting (\ref{eq:cov*}), setting $x=\tilde{y}$ in (\ref{eq:cdf.tilde.tn}) and Taylor expansion.

It remains to prove (\ref{eq:it.boot}) for the Studentized ${\cal I}_S(\alpha)$. 
We see by Taylor expanding the smooth function $H(\cdot)$ and the moment relations asserted in Lemmas~\ref{lem:mom*} and \ref{lem:mom**} that
$\hat\tau^2=\hat\sigma^2+O_p(n^{-1}+\ell n^{-3/2})$ and $\tau^{*2}=\sigma^{*2}+O_p(n^{-1}+\ell n^{-3/2})$.
Expanding $\tau^*$ about $\hat\sigma$ based on (\ref{eq:sigma*}), we have, for $\xi\in(0,1)$,
\[
J^*(z_\xi)=\prob^*\left((b\ell)^{1/2}[H(\bar{X}^*)-H(\expn^*\bar{X}^*)]-\Delta^*_n\leq\hat{w}\right)+O_p(\ell n^{-1}+k^{-2}),
\]
where $\Delta^*_n$ is defined as in (\ref{eq:calib}) and
\begin{eqnarray*}
\hat{w}&=&\hat\sigma z_\xi+(2\hat\sigma)^{-1}z_\xi\left\{
\sum_{r,s=1}^d{H}_r{H}_s\left[Q^{r,s}-P^{r,s}+(k^{-1}-\ell^{-1})\chi^{r,s}_{2,2}\right]\right.\\
&&\mbox{\ \ \ \ \ \ \ \ \ \ \ \ \ \ \ \ \ \ \ \ \ \ \ }\left.+\,2\sum_{r,s,t=1}^d{H}_r{H}_{st}\chi^{r,s}_{2,1}\left(k^{-1/2}Q^t-\ell^{-1/2}P^t\right)
\right\}.
\end{eqnarray*}
Noting (\ref{eq:yr5}) and (\ref{eq:calib2}), we have
\begin{eqnarray}
J^*(z_\xi)&=&G^*(\hat{w})+n^{-1/2}2^{-1}\hat\sigma^{-3}z_\xi^2\phi(z_\xi)\nonumber\\
&&\times\left\{\sum_{r,s,t=1}^d{H}_r
{H}_s{H}_t\sum_{i,j=-\infty}^\infty \gamma^{r,s,t}_{i,j}
+2\sum_{r,s,t,u=1}^d{H}_r
{H}_s{H}_{tu}\chi^{r,t}_{2,1}\chi^{s,u}_{2,1}
\right\}\nonumber\\
&&+\,O_p(\ell n^{-1}+n^{-1/2}k^{-1}L_n+(k\ell')^{-1}n^{-1/2}L_n^2).
\label{eq:jn*}
\end{eqnarray}
Recall the expression for $\hat\alpha=\alpha+\delta_n+B_n+O_p(\ell n^{-1}+k^{-2})$.
Putting $z_\xi=G^{*-1}(1-\hat\alpha)/\hat\sigma$ in (\ref{eq:jn*}), we verify that
$J^*(G^{*-1}(1-\hat\alpha)/\hat\sigma)=1-\alpha+O_p(\ell n^{-1}+k^{-2})$, so that
$\hat\tau J^{*-1}(1-\alpha)=G^{*-1}(1-\hat\alpha)+O_p(\ell n^{-1}+k^{-2})$. Thus 
${\cal I}_S(\alpha)$ is equivalent asymptotically to ${\cal I}_C(\alpha)$ up to $O_p\left\{n^{-1/2}(\ell n^{-1}+k^{-2})\right\}$, yielding for its
coverage probability the same expression as given by
(\ref{eq:cov*}) up to order $O(\ell n^{-1}+k^{-2})$. This completes the proof of part (ii).
\hfill\rule{0.5ex}{2ex}

\subsection{\label{app2}Other Studentizing approaches}
Under the smooth function model setting, Davison and Hall (1993) and G\"{o}tze and K\"{u}nsch (1996) suggest Studentizing the block bootstrap based on closed-form
expressions. Their constructions are similar to that of our ${\cal I}_S(\alpha)$, except that $\hat\tau$ and $\tau^*$ are replaced by closed-form expressions depending on 
partial derivatives $\{H_r\}$ of $H$. Specifically, Davison and Hall (1993) define 
$\hat\tau^2=\sum_{r,s=1}^d{H}_r(\bar{X}){H}_s(\bar{X})\hat\Sigma_{rs}$, where $\hat\Sigma_{rs}=n^{-1}\sum_{i=1}^n(X_i-\bar{X})^{(r)}(X_i-\bar{X})^{(s)}
+n^{-1}\sum_{j=1}^{\ell-1}\sum_{i=1}^{n-j}(X_i-\bar{X})^{(r)}(X_{i+j}-\bar{X})^{(s)}$, and $\tau^*$ analogously with ${\cal X}$ replaced by the block bootstrap series 
${\cal X}^*$ in the above definition of $\hat\tau$. G\"{o}tze and K\"{u}nsch's (1996) Studentizing factors have similar expressions except that they define $\hat\Sigma_{rs}
=\sum_{j=0}^{\ell-1}w_jn^{-1}\sum_{i=1}^{n-\ell}(X_i-\bar{X})^{(r)}(X_{i+j}-\bar{X})^{(s)}$, where $w_0=1$ and $w_j=2\{1-c(j/\ell)^2\}$ for $1\leq j\leq\ell-1$ and some $c>0$, and its bootstrap version
by $b^{-1}\sum_{j=1}^b\ell^{-1}\left\{\sum_{i=1}^\ell(X^*_{(j-1)\ell+i}-\bar{X}^*)^{(r)}\right\}\left\{\sum_{i=1}^\ell(X^*_{(j-1)\ell+i}-\bar{X}^*)^{(s)}\right\}$.

\newpage
\thispagestyle{empty}
\pagestyle{empty}
\begin{table}[h]
\begin{center}
{\footnotesize
\begin{tabular}{|l||rrrr|rrrr|}
\hline
& \multicolumn{4}{c|}{$n=500$}& \multicolumn{4}{c|}{$n=1000$} \\
nominal level $\alpha$ & 0.05 & 0.10 & 0.90 & 0.95 & 0.05 & 0.10 & 0.90 & 0.95\\ \hline\hline
\multicolumn{9}{|l|}{(a) ARCH(1) series}\\ \hline
${\cal I}(\alpha)$ & 0.053 & 0.099 & 0.897 & 0.943 & 0.034 & 0.110 & 0.903 & 0.939 \\
${\cal I}_C(\alpha)$ & 0.056 & 0.096 & 0.897 & 0.942 & 0.037 & 0.113 & 0.903 & 0.936 \\
${\cal I}_S(\alpha)$ & 0.052 & 0.097 & 0.898 & 0.944 & 0.034 & 0.109 & 0.901 & 0.937 \\
${\cal I}_{DH}(\alpha)$ & 0.053 & 0.100 & 0.899 & 0.944 & 0.033 & 0.106 & 0.902 & 0.939 \\
${\cal I}_{GK}(\alpha)$ & 0.052 & 0.102 & 0.899 & 0.941 & 0.036 & 0.107 & 0.902 & 0.935 \\ \hline\hline
\multicolumn{9}{|l|}{(b) MA(1) series}\\ \hline
${\cal I}(\alpha)$ & 0.059 & 0.088 & 0.904 & 0.948 & 0.050 & 0.104 & 0.899 & 0.952 \\
${\cal I}_C(\alpha)$ & 0.056 & 0.086 & 0.912 & 0.951 & 0.048 & 0.096 & 0.902 & 0.952 \\
${\cal I}_S(\alpha)$ & 0.053 & 0.085 & 0.914 & 0.954 & 0.044 & 0.098 & 0.904 & 0.952 \\
${\cal I}_{DH}(\alpha)$ & 0.052 & 0.087 & 0.912 & 0.955 & 0.048 & 0.097 & 0.902 & 0.951 \\
${\cal I}_{GK}(\alpha)$ & 0.053 & 0.087 & 0.908 & 0.953 & 0.046 & 0.098 & 0.900 & 0.954\\ \hline\hline
\multicolumn{9}{|l|}{(c) AR(1) series}\\ \hline
${\cal I}(\alpha)$ & 0.059 & 0.104 & 0.894 & 0.937 & 0.049 & 0.108 & 0.891 & 0.934 \\
${\cal I}_C(\alpha)$ & 0.045 & 0.096 & 0.902 & 0.941 & 0.043 & 0.104 & 0.899 & 0.939 \\
${\cal I}_S(\alpha)$ & 0.045 & 0.095 & 0.902 & 0.942 & 0.041 & 0.103 & 0.899 & 0.942 \\
${\cal I}_{DH}(\alpha)$ & 0.046 & 0.100 & 0.902 & 0.941 & 0.045 & 0.105 & 0.896 & 0.938 \\
${\cal I}_{GK}(\alpha)$ & 0.046 & 0.101 & 0.902 & 0.940 & 0.043 & 0.104 & 0.899 & 0.939\\ \hline
\end{tabular}}
\end{center}
\caption{\label{tab:mean}Mean example --- coverage probabilities of nominal level $\alpha$ upper confidence bounds for mean, approximated from 1,000 
independent series of length $n$.}
\end{table}

\newpage
\thispagestyle{empty}
\begin{table}[h]
\begin{center}
{\footnotesize
\begin{tabular}{|l||rrrr|rrrr|}
\hline
& \multicolumn{4}{c|}{$n=500$}& \multicolumn{4}{c|}{$n=1000$} \\
nominal level $\alpha$ & 0.05 & 0.10 & 0.90 & 0.95 & 0.05 & 0.10 & 0.90 & 0.95\\ \hline\hline
\multicolumn{9}{|l|}{(a) ARCH(1) series}\\ \hline
${\cal I}(\alpha)$ & 0.025 & 0.090 & 0.840 & 0.902 & 0.028 & 0.089 & 0.828 & 0.889 \\
${\cal I}_C(\alpha)$ & 0.053 & 0.111 & 0.887 & 0.943 & 0.054 & 0.109 & 0.869 & 0.926 \\
${\cal I}_S(\alpha)$ & 0.052 & 0.112 & 0.888 & 0.944 & 0.052 & 0.106 & 0.871 & 0.926 \\
${\cal I}_{DH}(\alpha)$ & 0.054 & 0.114 & 0.881 & 0.940 & 0.055 & 0.108 & 0.864 & 0.924 \\
${\cal I}_{GK}(\alpha)$ & 0.058 & 0.113 & 0.883 & 0.942 & 0.053 & 0.110 & 0.867 & 0.927 \\ \hline\hline
\multicolumn{9}{|l|}{(b) MA(1) series}\\ \hline
${\cal I}(\alpha)$ & 0.046 & 0.090 & 0.883 & 0.930 & 0.056 & 0.097 & 0.872 & 0.921 \\
${\cal I}_C(\alpha)$ & 0.059 & 0.099 & 0.906 & 0.946 & 0.064 & 0.105 & 0.884 & 0.936 \\
${\cal I}_S(\alpha)$ & 0.058 & 0.100 & 0.909 & 0.948 & 0.065 & 0.105 & 0.883 & 0.935 \\
${\cal I}_{DH}(\alpha)$ & 0.059 & 0.101 & 0.905 & 0.944 & 0.064 & 0.105 & 0.881 & 0.938 \\
${\cal I}_{GK}(\alpha)$ & 0.061 & 0.103 & 0.907 & 0.944 & 0.064 & 0.105 & 0.877 & 0.937\\ \hline\hline
\multicolumn{9}{|l|}{(c) AR(1) series}\\ \hline
${\cal I}(\alpha)$ & 0.042 & 0.091 & 0.885 & 0.928 & 0.047 & 0.097 & 0.863 & 0.916\\
${\cal I}_C(\alpha)$  & 0.053 & 0.106 & 0.903 & 0.950  & 0.059 & 0.107 & 0.881 & 0.936\\
${\cal I}_S(\alpha)$ & 0.052 & 0.104 & 0.902 & 0.953 & 0.058 & 0.110 & 0.880 & 0.932 \\
${\cal I}_{DH}(\alpha)$  & 0.054 & 0.104 & 0.899 & 0.952 & 0.057 & 0.109 & 0.883 & 0.930\\
${\cal I}_{GK}(\alpha)$  & 0.055 & 0.107 & 0.901 & 0.949 & 0.057 & 0.108 & 0.883 & 0.927\\ \hline
\end{tabular}}
\end{center}
\caption{\label{tab:var}Variance example --- coverage probabilities of nominal level $\alpha$ upper confidence bounds for variance, approximated from 1,000 
independent series of length $n$.}
\end{table}

\newpage
\thispagestyle{empty}
\begin{table}[h]
\begin{center}
{\footnotesize
\begin{tabular}{|l||rrrr|rrrr|}
\hline
& \multicolumn{4}{c|}{$n=500$}& \multicolumn{4}{c|}{$n=1000$} \\
nominal level $\alpha$ & 0.05 & 0.10 & 0.90 & 0.95 & 0.05 & 0.10 & 0.90 & 0.95\\ \hline\hline
\multicolumn{9}{|l|}{(a) ARCH(1) series}\\ \hline
${\cal I}(\alpha)$ & 0.063 & 0.103 & 0.878 & 0.930 & 0.056 & 0.098 & 0.888 & 0.930 \\
${\cal I}_C(\alpha)$ & 0.054 & 0.096 & 0.885 & 0.937 & 0.049 & 0.098 & 0.891 & 0.934 \\
${\cal I}_S(\alpha)$ & 0.054 & 0.099 & 0.884 & 0.936 & 0.052 & 0.098 & 0.890 & 0.933 \\
${\cal I}_{DH}(\alpha)$ & 0.057 & 0.100 & 0.883 & 0.934 & 0.058 & 0.100 & 0.888 & 0.930 \\
${\cal I}_{GK}(\alpha)$ & 0.049 & 0.095 & 0.878 & 0.934 & 0.054 & 0.094 & 0.887 & 0.935 \\ \hline\hline
\multicolumn{9}{|l|}{(b) MA(1) series}\\ \hline
${\cal I}(\alpha)$ & 0.056 & 0.095 & 0.905 & 0.952 & 0.041 & 0.084 & 0.888 & 0.947 \\
${\cal I}_C(\alpha)$ & 0.052 & 0.087 & 0.901 & 0.953 & 0.039 & 0.078 & 0.886& 0.944 \\
${\cal I}_S(\alpha)$ & 0.052 & 0.087 & 0.899 & 0.949 & 0.037 & 0.075 & 0.884 & 0.947 \\
${\cal I}_{DH}(\alpha)$ & 0.051 & 0.090 & 0.903 & 0.952 & 0.041 & 0.079 & 0.886 & 0.947 \\
${\cal I}_{GK}(\alpha)$ & 0.022 & 0.058 & 0.920 & 0.966 & 0.026 & 0.051 & 0.914 & 0.962\\ \hline\hline
\multicolumn{9}{|l|}{(c) AR(1) series}\\ \hline
${\cal I}(\alpha)$ & 0.067 & 0.110 & 0.882 & 0.941 & 0.045 & 0.101 & 0.873 & 0.936\\
${\cal I}_C(\alpha)$  & 0.057 & 0.100 & 0.880 & 0.941  & 0.042 & 0.093 & 0.877 & 0.935\\
${\cal I}_S(\alpha)$ & 0.055 & 0.101 & 0.881 & 0.937 & 0.039 & 0.091 & 0.878 & 0.937 \\
${\cal I}_{DH}(\alpha)$  & 0.060 & 0.103 & 0.882 & 0.946 & 0.044 & 0.095 & 0.873 & 0.941\\
${\cal I}_{GK}(\alpha)$  & 0.025 & 0.065 & 0.897 & 0.959 & 0.023 & 0.054 & 0.899 & 0.955\\ \hline
\end{tabular}}
\end{center}
\caption{\label{tab:corr}Autocorrelation example --- coverage probabilities of nominal level $\alpha$ upper confidence bounds for lag 1 autocorrelation, approximated from 1,000 
independent series of length $n$.}
\end{table}

\end{document}